\documentclass[oneside,a4paper,reqno]{amsart}
\usepackage{amssymb}

\theoremstyle{remark}

\begin{document}

\begin{titlepage}
\begin{center}
\bfseries  	FACTS, VALUES AND QUANTA
\end{center}
\vspace{1 cm}
\begin{center} D M APPLEBY
\end{center}
\begin{center} Department of Physics, Queen Mary
University of London,  Mile End Rd, London E1 4NS,
UK
 \end{center}
\vspace{0.5 cm}
\begin{center}
  (E-mail:  D.M.Appleby@qmul.ac.uk)
\end{center}
\vspace{0.75 cm}
\vspace{1.25 cm}
\begin{center}
\vspace{0.35 cm}
\parbox{12 cm }{ 
  Quantum mechanics is a fundamentally probabilistic theory (at least so
far as the empirical predictions are concerned).  It follows that,
if one wants to properly understand  quantum mechanics, it is essential
to clearly understand the meaning of probability statements.  The
interpretation of probability has excited nearly as much philosophical
controversy as the interpretation of quantum mechanics. 
$20^{\mathrm{th}}$ century physicists have mostly adopted a frequentist
conception.  In this paper it is argued that we ought, instead, to adopt
a logical or Bayesian conception.  The paper includes a comparison
of the orthodox and Bayesian theories of statistical inference. It
concludes with a few remarks concerning the implications for the concept
of physical reality. 
   }
\end{center}
\end{titlepage}
\section{Introduction}
This paper is about probability.  It was originally stimulated by some
conversations with Chris
Fuchs~\cite{FuchsEtAl,FuchsEtAlB,FuchsEtAlC,FuchsSam1,FuchsSam2}
concerning the foundations of quantum mechanics.  These conversations had
a major impact on my thinking:  they made me wonder if there might, after
all, be something to be said for the Copenhagen interpretation.

I did at first doubt whether it is appropriate to dedicate these
remarks to the memory of Jim Cushing, who expended much energy on the
task of challenging the Copenhagen hegemony~\cite{CushingA}.  But after
reflection  I decided that it is very appropriate.    Jim has spent a
great deal of  time and effort studying Copenhagen ideas:  far more than
I have myself.  I feel that he would have to be interested at least in
the question.  What he would think of my (tentative) answer  is,
unfortunately, impossible now to know.  But what I do know is that there
is no one whose opinion I would more eagerly have sought.

Jim was, before he was anything else, a friend of reason. He had no
objection  to someone who  adopts  Copenhagen
assumptions provisionally, in a spirit of free enquiry,  to see where
they lead. His objection was to dogmatic Copenhagenism:  the insistence
that Bohr and his colleagues   settled the question, completely, once and
for all, and that there is nothing more of any interest to be said on the
subject.

Nowadays this attitude would be quite unusual. Even those who are
sympathetic to Copenhagen ideas would mostly  agree that there
are many remaining obscurities.  There are, besides, numerous other
interpretations on offer, all well represented in the pages of the
leading journals.  However, it was not always so.   The climate which
existed $30$ years ago is well illustrated by the editorial
comment~\cite{BallEd} in which the editor of
\emph{Reviews of Modern Physics} defends his decision to publish
Ballentine~\cite{BallPaper}.  It can be seen from this comment that there
was then a  substantial body of opinion which held that Einstein's
ideas regarding the interpretation of quantum mechanics should not be
discussed in the scientific literature. It is this refusal even to
consider the question which Bell~\cite{Bell} (p.~160) aptly
describes as Copenhagen ``complacency''.

Fortunately those days are now over.  We seem to have finally got back to
the situation which existed in the  1920's,  where 
the Copenhagen case has to be argued.  We owe this happy state of affairs 
to the efforts of Jim Cushing and others who have struggled to convince
the physics community that Bohr's word should not be taken as final. I
think that even those who favour the  Copenhagen approach  owe him a
debt.  An atmosphere of bland orthodoxy, in which stimulating discussion
is not to be had, is unlikely to foster new ideas. 

Fuchs~\cite{FuchsSam1} (p.~173) advocates the Copenhagen interpretation
in the following terms: 
\begin{quote}
``I have this `madly optimistic' (Mermin
called it) feeling that Bohrian-Paulian 
ideas will lead us to the next stage of physics. That is, that
thinking about quantum foundations from 
their point of view will be the \emph{beginning} of a new path,
not the end of an old one.''
\end{quote}
Proposed like that, not as a piece of  dogmatism, but as an
invitation to serious thought, the Copenhagen interpretation, to my mind,
suddenly becomes interesting.

This does not mean that I find the Copenhagen interpretation satisfactory,
as it  stands now.  Bell~\cite{Bell} (pp.~173--4)
argues that the Copenhagen interpretation is ``unprofessionally
vague and ambiguous''. I think he is right.  I also think he is right
to complain that quantum mechanics, when interpreted in traditional
Copenhagen terms, seems to be ``exclusively concerned with `results of
measurement' and [seems to have] nothing to say about anything else''.   
I share Bell's conviction that the aim of  physics is to understand
nature, and that counting detector ``clicks'' is not intrinsically any
more interesting than counting beans.  If prediction and control were my
aim in life I would have become an engineer, not a physicist.  

Fuchs has not caused  me to see clarity where I formerly saw obscurity,
or realism where I formerly saw positivism.  What he has done is to
make me wonder if it might be possible to constuct a greatly improved
version of the Copenhagen interpretation, to which these objections
would not apply.  If I am asked to accept Bohr as the
authoritative voice of final truth, then I cannot assent.  But if
his writings are approached in a more flexible spirit, as a source
of insights which are not the less seminal for being obscure, they
suggest some interesting questions.  I do not know if this line of
thought will be fruitful.   But I feel it is worth pursuing.   I
also feel  that Jim Cushing would  consider it worth pursuing.

I will make a few comments concerning the question of realism in
Sections~\ref{sec:Propensity} and~\ref{sec:conclusion}.  However in this
paper I will mainly be concerned with probability.   Fuchs and his
co-workers have made a number of significant conceptual innovations (they
are, I believe, the first members of the Copenhagen tendency since the
1930's who, not content merely  to reiterate pieces of received
orthodoxy, seriously try to advance the theory at a basic conceptual
level).  One such innovation is their analysis of probability, and its
relevance to the interpretation of quantum mechanics (also see  
important work by Hardy~\cite{HardyA,HardyB},
Pitowsky~\cite{Pitowsky} and Perey~\cite{Perey}).

Quantum mechanics is a fundamentally probabilistic theory.  Of course,
probability theory plays an essential role in classical physics too. 
However, in classical physics the uncertainties can, in principle, be
made arbitrarily small.  In quantum physics they are 
ineluctable\footnote{
  Or so it now seems.  This statement might need to be modified
  if Valentini's~\cite{ValentiniA,ValentiniB} ideas were empirically
confirmed. }.
So it is  not unreasonable to suggest that, to properly
understand quantum mechanics, we need first to straighten out our
ideas regarding probability, and its physical significance.

Whereas the interpretation of quantum mechanics has only been puzzling us
for
$\sim 75$ years, the interpretation of probability has been doing so
for more than  $ 300$ years~\cite{HackingA,Daston}.  
Poincar\'{e}~\cite{Poincare} (p.~186) described  probability as ``an
obscure instinct''.  In the century that has elapsed since
then philosophers  have
worked hard to lessen the obscurity.  However, the result has not been to
arrive at any consensus.  Instead, we have a number of competing 
schools (for an overview   see  
Gillies~\cite{GilliesA}, von Plato~\cite{vonPlato},
 Sklar~\cite{SklarA,SklarB} and Guttman~\cite{Guttman}).

The majority of $20^{\mathrm{th}}$ century physicists subscribed to a
frequency interpretation of probability.  But in the
$19^{\mathrm{th}}$ century a very different view was widely held.  It is
exemplified by the following remark of Maxwell's:
\begin{quote}
They say that Understanding ought to work by the rules of right
reason.  These rules are, or ought to be, contained in Logic; but the
actual science of logic is conversant at present only with things either
certain, impossible, or entirely doubtful, none of which (fortunately) we
have to reason on.  Therefore the true logic for this world is the
calculus of Probabilities, which takes account of the magnitude of the
probability which is, or ought to be, in a reasonable man's mind.
[J. Clerk Maxwell, quoted Jeffreys~\cite{JeffreysB}, p.1]
\end{quote}
Maxwell here espouses what I am going to call an epistemic view of
probability.  As Maxwell sees it a probability statement has a
normative, or logical significance.   It does not directly assert a fact
about the way things are in the world.  Instead it regulates our
expectations concerning the world. 
Under the influence of Bayes and Laplace~\cite{Laplace} this
way of looking at probability was common for a large part of the
$19^{\mathrm{th}}$ century. 

Suppose that Maxwell's demon is interested in the question, whether a
(classical) molecule is going to hit its shutter during the
interval $(t,t+\Delta t)$ at some specified  time $t$ in the future. 
Suppose that to begin with the demon only knows the temperature, pressure
and volume of the gas.  On that basis the demon calculates that the
probability $=$ (say) $0.001$.   Suppose, however, that the demon then
acquires detailed information regarding the positions and velocities of
all the gas molecules at some time $< t$; and suppose that on the basis
of that new information, and Newtonian mechanics, it calculates that a
molecule is certain to arrive at its shutter during the interval
$(t,t+\Delta t)$.  Then the probability changes discontinuously from
$0.001$ to $1$ (the probability distribution might be said to
``collapse'').  But this discontinuous change in the probability does not
reflect any change in the state of the gas.  All that has changed is the
state of the demon's knowledge.  That is what is meant by calling the
probability epistemic.

However, towards the end of Maxwell's life the epistemic view began to
go out of fashion, and  throughout the
$20^{\mathrm{th}}$ century it was very 
unfashionable
indeed\footnote{
  I think that must be fair as a general statement.  Nevertheless, 
  the epistemic view has
 continued to engage the interest of a very active minority.
$20^{\mathrm{th}}$ century advocates  include
Keynes~\cite{Keynes}, Jeffreys~\cite{JeffreysB,JeffreysC},
Carnap~\cite{Carnap},  Lewis~\cite{Lewis}, 
 de Finetti~\cite{Fin2,Fin1}, Savage~\cite{Savage}, Bernardo and
Smith~\cite{Bernardo}, Jaynes~\cite{Jaynes}, Howson and
Urbach~\cite{Howson} and
Earman~\cite{Earman} (this list is not intended to be
complete).  The list might be  sub-divided into 
those~\cite{Keynes,JeffreysB,JeffreysC,Carnap} who think
that probability involves a new kind of non-deductive logic and
those~\cite{Fin2,Fin1,Savage,Bernardo,Jaynes,Howson,Earman} who
take a Bayesian approach.  However, this classification  is 
somewhat arbitrary.  Jeffreys, in particular, straddles the two
categories. Jaynes should be singled out for special mention
because he is primarily concerned with applications to physics.
 }.  
In physics
it was largely replaced with the frequentist
conception~\cite{Venn,Mises,MisesB,Reich,PopperB,Fraassen},
according to which a probability can be identified with a relative
frequency in some suitably defined ensemble.  The attraction of
this view is that, by removing all reference to the knowledge
and/or beliefs of some cognitive agent (human or otherwise), it
promises to make the concept of probability purely objective.

Fuchs \emph{et al} consider that this was a mistake, and that we need to
go back to an epistemic conception.  When I first became acquainted with
their ideas I resisted this suggestion .  However, the more I
have thought about it, the more I have become convinced that they have to
be right. This paper is the fruit of those cogitations.

Fuchs
\emph{et al}  subscribe to the epistemic theory
proposed  by de Finetti~\cite{Fin2,Fin1,Savage,Bernardo} (also see
Ramsey~\cite{Ramsey}).   I should say that I do not entirely agree with
them about that.   Although I fully acknowledge the depth and importance
of de Finetti's insights, it seems to me that he misses some essential
points.  My  feeling is that a completely satisfactory theoretical
account  has yet to be
formulated.  However, that  has no bearing on my argument here.  In this
paper I am simply concerned to argue that the epistemic conception is,
\emph{in one form or another},  unavoidable. 

Much of the paper concerns the theory of statistical
inference, which has a crucial bearing on the question. An essential part
of the frequentist position is that probabilities are, not only objective,
but also in some sense observable.   If the proposal was instead that
probabilities, though purely objective, are empirically unknowable---if,
in other words, probabilities were conceived as hidden variables---then
the frequentist idea would lose most, if not all, of its attractiveness.

The method of inference originally proposed by Bayes and Laplace is
unsatisfactory from the frequentist point of view because the inferred
probability distribution depends, not only on the empirical data, but
also on a prior probabilistic assumption.   Suppose, for example, that a
coin comes up heads on $500$ out of $1000$ successive tosses.  Then, with
the appropriate choice of prior assumption, the inferred distribution can
be strongly concentrated in the vicinity of \emph{any} probability in the
interval $[0,1]$.   This is not a problem if one looks at it from the
epistemic point of view which Maxwell describes in the passage I quoted
above.  However, from a frequentist point of view it is a very serious
problem.  For the frequentist programme to work a statistical
inference needs to be something like a measurement:
it must be possible to read the inferred disribution 
directly off from the data, without the assistance of any prior
assumption.

In the opening decades of the $20^{\mathrm{th}}$ century
Fisher~\cite{Fisher}, Neyman and Pearson~\cite{Neyman} and others
accordingly developed a new statistical methodology, in conscious
opposition to the ideas of Bayes and Laplace. This is the approach based
on confidence intervals and hypothesis tests which, for want of a
better 
term\footnote{
  I prefer not to use the term ``frequentist methodology''. Although the
question of 
  inferential methodology and the question of interpretation are
intimately related, I think it is better to keep them terminologically
distinct.
  Jaynes~\cite{JaynesB} uses the term ``orthodox statistics''.  Since 
   then use of Bayesian techniques has become
 much more common.  However, even if the description ``orthodox
methodology'' is no longer  accurate, I feel it may still serve as a
term of convenience. 
},   
I am going to call the orthodox methodology.    The
attraction of the orthodox methodology is that, unlike the Bayesian
methodology, it seems to make statistics purely empirical, and purely
objective.  One of the main conclusions of this paper is that it fails in
that purpose.   Not only is it, as de Finetti~\cite{Fin1} (vol.~2,
p.~245) says, \emph{ad hoc}.  It is no  less dependent on prior
probabilistic assumptions than the Bayesian methodology.

Hume~\cite{HumeA} (p.~469) famously argued that one cannot validly
infer an ``ought'' from an ``is''.  A similar principle applies to
probability statements:   one cannot validly infer a ``probable''
from an ``is''. This principle is  closely related to the
conclusion to Hume's
  argument
  for inductive scepticism~\cite{HumeA,HumeB}.  It means that
probability judgments are not purely empirical.  It also means that
a probability statement cannot be identified with a fact about the
world, as it exists independently of us.

It was easy for Maxwell to accept an epistemic interpretation of
probability because he was thinking in terms of a world populated by
classical atoms and fields, whose objective reality was not in doubt. 
But if one translates the idea to a quantum mechanical context, and
suggests that the quantum state must be interpreted epistemically, then
the concept becomes very disturbing.  It certainly becomes
disturbing to me (although I find Fuchs's ideas stimulating, they
also worry me). On the face of it, taking an epistemic view of the
state vector amounts to giving up on the idea of physical reality
altogether.

I remain very uncertain about this.  An epistemic
interpretation of the state vector is, it seems to me, impossible
to reconcile  with Einsteinian realism: the proposition that ``the
programmatic aim of all physics [is] the complete description  of
any (individual) real situation (as it supposedly exists
irrespective of any act of observation or
substantiation)''~\cite{Einstein} (p.~667).  However,  I feel it
may be consistent with a much more subtle and interesting kind of
realism, which is obscurely intimated in the writings of Bohr, but
which has yet to be properly articulated.  I will touch on this  in
Section~\ref{sec:conclusion}.

The plan of the paper is as follows.  
Sections~\ref{sec:frequentism1} and~\ref{sec:frequentism2}
concern the frequency interpretation.   Section~\ref{sec:Propensity}
concerns the idea of a propensity. 
Sections~\ref{sec:Bayes}--\ref{sec:noprobable} are the core of the paper
and are concerned with the theory of statistical inference. 
Section~\ref{sec:predictive} (also Section~\ref{sec:frequentism2})
concerns the idea of impossibility FAPP (``for all practical purposes''). 
Section~\ref{sec:frequentism3} contains some final criticisms of the
frequentist view, based on the argument in
Sections~\ref{sec:Bayes}--\ref{sec:predictive}. 
Section~\ref{sec:normative} concerns the epistemic view. 
Section~\ref{sec:conclusion} concerns
the question of physical realism.

\section{Frequentism:  Infinite Ensembles}
\label{sec:frequentism1}

Physicists naturally tend to favour a frequency interpretation of
probability.  According to this conception the proposition ``the
probability of  this coin coming up heads $=1/2$'' is a 
straightforward
factual assertion about the number of heads in a long or infinite
sequence of tosses.   The idea is attractive because it promises
to give  probability statements a purely objective significance. 
It is originally due to Venn~\cite{Venn}.  The best known
$20^{\mathrm{th}}$ century proponents are 
von Mises~\cite{Mises,MisesB}, Reichenbach~\cite{Reich},
Popper~\cite{PopperB} (in the first part of his career) and van
Fraassen~\cite{Fraassen}.

H\'{a}jek~\cite{Hajek} (endnote~2)  says that among
philosophers frequentism is now largely confined to the
``closet''.  But  he also remarks that it continues to pervade much
scientific thinking on the subject.  It is not difficult to
understand why.  Once acknowledge that frequentism is untenable,
and one is forced to re-assess some of the most basic assumptions
of physical theory.  Philosophers can afford to take a
comparatively relaxed attitude to this necessity.  Physicists have
more of an investment.

If one wants to maintain that
 there is  an effective\footnote{
  \label{fnte:infinitefrequentism}
 The equivalence cannot be strict even in the case of 
   infinite sequences. It is true
  that if the probability of heads
   $=0.5$ then, by the strong law of large
numbers~\cite{Feller} (vol.~1, pp.~203--4),  the set of
  sequences 
 	in which the limiting relative
  frequency of heads either does not exist, or exists
 but $\neq 0.5$, has
  measure zero (relative to the
   product measure
  on
   the space 
  of infinite sequences).  But
that is not
 equivalent to the proposition  that such sequences are impossible.
}
 logical equivalence
between a statement about the probability of a coin 
coming up  heads, and
a statement about the actual frequency of heads in a sequence 
of tosses,
then it is clearly essential that the sequence be 
infinite.  
 There is a
problem with this, however:  for it is (to say the least) doubtful
whether a coin physically could be tossed an infinite number of
times.   Leaving
aside the problems of corrosion, and mechanical wear, 
there is the problem
that we expect the sun eventually to become a red giant.  
Supposing the
coin to survive that
 vicissitude one then has the problem that the lifetime of the
proton, not to mention the universe, may be finite.

von Mises's approach is to define the probability
counter-factually, as the limiting relative frequency which
\emph{would} be obtained \emph{if} the coin \emph{were} tossed an
infinite number of times (see, for example, von Mises~\cite{Mises},
p.15).  This might be acceptable if the aim was only to provide a
convenient way to think about probabilities (though I would
question whether it really is all that convenient).  But, as
Jeffrey~\cite{Jeffrey} says, it is clearly unacceptable if the aim
is to identify a probability with an actually existent physical
quantity, out there in the world.  If probabilities are to
objectively exist, then it is essential that the sequences in terms
of which they are defined should  objectively exist.

The universe may have finite $4$-volume.  In that case, on an
infinite frequentist definition, there would not be any
probabilities.  At least, there would not be any probabilities of
directly observable events, having non-zero 
spatio-temporal extension.

However, that is not the only difficulty with the infinite
frequentist idea.  Suppose we grant, for the sake of argument,
that the universe has infinite
$4$-volume.  A theory which makes the
 probabilities of events now, here in the
Milky Way, critically dependent on events  more than
$10^{10^{100}}$ years in the future, or more than 
$10^{10^{100}}$ light years distant, cannot be
considered empirically relevant.

 Let $S$ be some suitably enormous, but still finite region of
space-time containing ourselves.  Let $E$ be the finite ensemble of
$\mathstrut^{226} \mathrm{Ra}$ nuclei whose world-lines intersect
$S$, and let 
$E_\infty$ be the ensemble consisting of all
$\mathstrut^{226}
\mathrm{Ra}$ nuclei  in the
universe (assumed infinite).  Let
$f$ be the proportion of nuclei in
$E$ which decay  in proper time $1602$ years, and let $f_\infty$ be
 the
proportion
 in 
$E_\infty$ which do 
so (as defined by some appropriate limiting procedure).  Suppose it
should happen that $f=1/2$, but $f_\infty = 1/10$. Then, on an
infinite frequentist definition, we are obliged to say that the true
decay-probability  is
$1/10$.  It would, however, seem  more natural to say that
the half-life  of $\mathstrut^{226} \mathrm{Ra}$ is $1602$ years 
in our
part of the universe, but  that it takes different values elsewhere.

It is natural to ask if the parameters defining the standard model
depend on spatio-temporal position.  If so one would
  expect
half-lives typically
 to depend on spatio-temporal position.  But on an infinite
frequentist definition that suggestion is not even meaningful.

Suppose that  a roulette wheel in London is spun
$37,000$ times, and  each number comes up approximately
$1,000$ times.  Clearly, this has no implications for the fairness
of a  different, completely unrelated roulette wheel in Rio de
Janeiro.  By the same token, it has no implications for the
fairness of the  roulette wheel in London in $1000$ years time
(supposing it  still to exist).  

Similarly, a determination of the half-life of $\mathstrut^{226}
\mathrm{Ra}$ on the Earth now has, in itself, without extra
assumptions, no implications for what the half-life of 
$\mathstrut^{226}
\mathrm{Ra}$ is at a point $10^{10^{100}}$
light-years distant,  or will be at a time 
$10^{10^{100}}$ years in the future.

Measuring the relative frequency in a finite ensemble in London
in the year $2004\,\text{\textsc{ad}}$ does, in itself, without
additional assumptions, tell one nothing about the limiting
relative frequency in some (purely hypothetical, empirically
inaccessible) embedding ensemble extending over an infinitely large
spatio-temporal region.  Similarly, if
\emph{per impossibile} one knew the ``true'' probability in the
sense of an infinite frequentist definition, this would tell one 
nothing about the relative frequencies to be expected in the finite
ensembles of actual interest.  This is, in essence, just the point
of Hume's argument for inductive scepticism~\cite{HumeA,HumeB}.

Probabilities in the sense of an infinite frequentist definition
may perhaps exist.  But  they have nothing to do with the
probabilities which we infer from our observations, and on which we
base our practical decisions.  They are empirically and practically
irrelevant.  

Suppose we discovered that the universe does have finite
$4$-volume.  This would not affect practical probabilistic
reasoning in any way.  Tabulations of nuclear half-lives 
would not suddenly be rendered meaningless.

\section{Frequentism:  Finite Ensembles}
\label{sec:frequentism2}
If one wants to give a frequentist definition of the probabilities
which occur in empirical reasoning, then the definition had better
be in terms of
\emph{finite} ensembles.  

The shift to finite ensembles necessitates a significant weakening
of the frequentist position.  The proposition ``the probability of
heads  $=0.5$'' is 
consistent\footnote{
  As I remarked in footnote~\ref{fnte:infinitefrequentism} it is 
  logically consistent in the infinite case also.  However, in the
 infinite
 case there is zero probability of obtaining a limiting relative
 frequency $\neq 0.5$.
}
 with \emph{any}  sequence
of outcomes in a run of (say) $1000$ tosses.    It is admittedly
very unlikely that  the coin will come up heads on each of $1000$
tosses.  But the probability of this happening is $>0$.

The usual response to this difficulty is to argue that very small
probabilities count as practical
impossibilities.  On those
grounds it is
suggested that the proposition ``the probability of
heads 
$=0.5$'', though not strictly equivalent, is \emph{for all practical
purposes} 
equivalent
to the proposition ``the relative frequency of
heads will be extremely close to $0.5$ in a sufficiently long
sequence of independent tosses''.

In the philosophical literature an even weaker position is often
advocated.  According to  Popper~\cite{PopperB}  (also see
Gillies~\cite{GilliesA,GilliesB} and, for a critical discussion,
Howson and Urbach~\cite{Howson}) the proposition 
``the probability of
heads 
$=0.5$'' is not confirmed by the outcome of any finite sequence of
tosses.  It is, however,  practically falsified if the sequence is
sufficiently long, and if the relative frequency of heads is
sufficiently different from $0.5$.  
Gillies~\cite{GilliesA} (p.~147) argues that this 
coincides with the
principle on which statistical hypothesis testing is based.

There is a problem with this line of thought.  Even a
probability of $2^{-1000}$, though it can  be neglected
for \emph{many} practical purposes, cannot  be neglected
for all. 

Suppose a coin comes up heads on each of $24$ successive tosses. 
Then, if one judges by the standards usual in statistical
hypothesis testing, one will reject the hypothesis that the coin
is fair.  

It is tempting to see this reasoning as a ``for all practical
purposes'' or FAPP version of an argument in formal logic.
Suppose
\begin{equation}
P \wedge Q \Rightarrow \bar{R}
\end{equation}
(where $\bar{R}$ signifies ``not $R$'').   Suppose it is known that
$P$ and $R$ are both true.  Then it follows that $Q$ is false. 

The argument concerning the $24$ coin tosses looks, superficially,
like a ``wobbly'' version of this, in which the logical rigours
have been slightly relaxed.  
Let $P=$  ``the coin is tossed $24$ times, the tosses being
independent'',
$Q=$  ``the coin is fair'' and
$R=$  ``$24$ heads obtained''.   Then it is tempting to think
\begin{equation}
P \wedge Q \Rightarrow
\tilde{R}
\label{eq:WobblyImplyA}
\end{equation} 
(where
$\tilde{R}$ signifies ``FAPP not $R$'').  
We know (or  assume) that $P$ and $R$ are both true.  So 
we conclude that
$Q$ is   false FAPP.

However, that badly misrepresents the real logic of the argument. 
Instead of $24$ coin tosses consider a lottery with 
$2^{24}$ tickets (this is about the number of tickets in the
British national lottery). 
Let $P=$  ``Alice buys one ticket'', $Q=$  ``the lottery is fair''
and $R=$ ``Alice wins''.  Then, if one accepts the
reasoning in the last paragraph, one must apparently  accept that
here too
\begin{equation}
P \wedge Q \Rightarrow \tilde{R}.
\label{eq:WobblyImplyB}
\end{equation}
 Suppose, now, that
Alice  does buy one ticket, and the ticket does win.  Then, if the
reasoning in the last paragraph is valid, it would follow that here
too
$Q$ is  false FAPP.  However, that
conclusion is clearly not justified.  One cannot reasonably infer
that a lottery is unfair just on the grounds that somebody wins it.

The fallacy in this argument is the assumption leading to
Eq.~(\ref{eq:WobblyImplyB}):  that if $R$ is highly improbable,
then
$R$ is effectively impossible.  This  assumption is justified in
some situations, but not in others.  The problem is to decide
exactly when it is justified.  

Let us note that the problem has nothing essentially to do with the
exact size of the probabilities. The  argument based
on  Eq.~(\ref{eq:WobblyImplyB}) would  still be invalid if Alice had
won a cosmic lottery having $10^{100}$  tickets.  The
microstate of the air in the room where I am now writing is even
more improbable:  but it still happened.

In considering this question one needs to distinguish two different
kinds of probabilistic argument, which I will call predictive and
retrodictive.

A predictive
argument is one in which conclusions are drawn regarding an unknown 
event (``predictive'' because, although the event may be in the
past, the discovery as to whether it happened lies in the future). 
For instance, Alice judges that she is very unlikely to win the
lottery.  So she works on the assumption that she will still 
be in salaried employment next month. 

A retrodictive argument moves in the reverse direction.  After the
data has come in, one uses the information to revise, or update
one's original probability assignment.  For instance, after
observing a run of 24 heads, one rejects the hypothesis that the
coin is fair.

 The two examples discussed above---the 24 coin tosses, and
AliceÕs lottery ticket---seem to differ very little so far as the
predictive aspect is concerned (though I will argue in
Section~\ref{sec:frequentism3} that the concept of FAPP
impossibility involves some subtleties even in the predictive
case).  On the other hand there seems to be a major difference from
the retrodictive point of view.  It is on that that I now want to
focus.

The first response of a convinced frequentist might be 
that the difference is due to  the fact that, whereas in
the first case one has an ensemble of $24$  events, in the
second case one is dealing with a singular event.  However, that
cannot be correct.  It is easy to think of situations where one 
\emph{can} validly draw retrodictive conclusions from singular
events.

For instance, the probability of a single $\mathstrut^{226}
\mathrm{Ra}$ nucleus decaying in the next hour is $5\times
10^{-8}$---about the same as the chance of Alice winning the
lottery.  Suppose that a single nucleus, initially believed to be 
$\mathstrut^{226} \mathrm{Ra}$, and held in a trap, actually does
decay in less than $1$ hour.  This might well cause one to doubt
whether it really was a $\mathstrut^{226}
\mathrm{Ra}$ nucleus---whether the decay probability really was
$5\times 10^{-8}$. 

What is the difference between this example and the example of
Alice's lottery ticket?  In both cases one has a singular
event, initially judged to be very improbable.  In the second case
the  occurrence of this event leads us to question the initial
probability assignment whereas in the first it does not.  Why is
that?  What is the underlying logical principle on which the
decision depends?

I am going to argue (here, and in
Sections~\ref{sec:Bayes}--\ref{sec:noprobable}) that the difference
has to do with our background assumptions.  We know that it is easy
to mis-identify a single nucleus and so, when the supposed
$\mathstrut^{226}
\mathrm{Ra}$ nucleus decays much sooner than expected,  that
naturally increases our readiness to embrace the alternative
hypothesis, that the experimenter made a mistake.  On the other
hand, the suggestion that the lottery might have been biased
\emph{specifically} in Alice's favour (as opposed to one of
the other $2^{24}$ ticket holders) initially strikes us as
comparatively  implausible.

One's judgment regarding the lottery might be different if one knew
more about it.  Suppose, for instance, one knew that Bob, who
runs the lottery, is Alice's best friend.  Then one might see the
event, that Alice wins the lottery, as cause for suspicion.

Suppose that Alice's winning ticket is number 5,592,405.  This
would not usually be seen as suspicious.  Suppose, however,  one
noticed that  in base $4$ her ticket number is   111111111111, and
suppose one then discovered that the lottery outcome is generated
by tossing a tetrahedral die 12 times.  Then the fact that this was
the winning number might seem very suspicious.

We get some further insight into the role of background assumptions
from a modification of Goodman's~\cite{Good} well-known ``grue''
argument.  Goodman uses this idea to analyze inductive reasoning. 
However, a  related argument applies to retrodictive probabilistic
reasoning (which can be seen as a generalized form of inductive
reasoning).

Goodman defines  ``grue'' to be the property ordinarily called
``green''  before a certain time $t$, and the property ordinarily
called  ``blue'' afterwards.  Before  $t$ every observed emerald
has been grue (which is to say green).   A naive inductive argument
would then suggest that observed emeralds will also be grue
(which is to say blue) after time
$t$.

To apply this idea in the case of interest here, choose some 
infinite sequence $x=(x_1,x_2,\dots)$ of $0$'s and $1$'s.  For the
sake of definiteness, let us take
$x$ to be the digits in the  binary expansion of the fractional
part of
$\pi$ (so $x=(1, 1, 0, 0, 1, 0, 0, 1, 0,\dots)$).  On the
$n^{\mathrm{th}}$ toss
 define ``heils'' to be the event ``heads'' if $x_n=1$ and
``tails'' if
$x_n=0$.  Define ``taads'' to be the event ``not heils''. 

Suppose now that in 24 tosses we obtain 24 heils and no taads.  In
conventional terms this is the sequence HHTTHTTHTTTTHHHHHHTHHTHT
consisting of 13 heads and 11 tails, and it would not usually be
seen as significant.   But if retrodictive inferences worked in
the way that Popper and Gillies think it would have to be seen as
highly significant.  If the hypothesis that the coin is fair is FAPP
falsified by $24$ heads in succession, then it must, on their
principles, also be FAPP falsified by $24$ heils in succession.

The reason we would not normally see $24$ heils in succession as
grounds for doubting  a coin's fairness has, I believe, to do with
our background beliefs. We attach significance to $24$ heads in
succession because we can  envisage a physical mechanism by which
the coin plausibly might be biased in favour of heads.  By
contrast, a bias in favour of heils seems, on physical grounds,
comparatively implausible.

Our assessment 
might be different if our background knowledge was greater. 
Suppose, for instance, we knew that the coin was ferromagnetic, and
suppose we also knew that there was a powerful electromagnet in the
vicinity which was switched on when
$x_n=0$, and off when $x_n=1$.  In that case we might consider $24$
heils in succession to be no less significant than $24$ heads in
succession. 

Probabilistic reasoning is, in short, very sensitive to
context. I will develop this point in
Section~\ref{sec:Bayes}--\ref{sec:predictive}.  I will resume my
discussion of the frequentist idea in
Section~\ref{sec:frequentism3}.

\section{Propensities}
\label{sec:Propensity}
Although Popper began as a frequentist he later switched to a
propensity interpretation~\cite{PopperC,PopperD,PopperE}.  A
substantial part of the philosophical community has followed him in
this (see Gillies~\cite{GilliesA,GilliesB}  and
references cited therein).

 Popper saw this development  as an evolution in his thought
(albeit an important evolution),  not a clean break with the past. 
Furthermore, it is a step which von Mises to some extent
anticipated, as Howson and Urbach~\cite{Howson} (p.~221) remark.  
According to von Mises~\cite{Mises} (p.14)
\begin{quote}
 ``The probability of a $6$ is a physical property of a given die
and is a property analogous to its mass, specific heat, or
electrical resistance.  Similarly, for a given \emph{pair of dice}
(including of course the total setup) the probability of a `double
6' is a characteristic property, a physical constant belonging to
the experiment as  a whole and comparable with all its other
physical properties. The theory  of probability is only concerned
with relations existing between physical quantities of this kind. ''
\end{quote} 
There are, of course, some important differences between Popper and
von Mises.  In particular, Popper admits  objective
single-case probabilities.  Gillies~\cite{GilliesA} (p.~114),
however, argues that this is  not essential to the
propensity concept.  Furthermore, the fact that von Mises defines
probabilities counter-factually (see Section~\ref{sec:frequentism1})
shows that he is really thinking of them as 
dispositional properties (defined contextually, relative to the
``experiment as a whole'', as he puts it in the above passage). In
short, it seems to me that, although von Mises is usually described
as a frequentist, he is in fact a propensity theorist.

I doubt whether there can ever really have been   a \emph{pure}
frequency theorist:  \emph{i.e.}\ a frequentist who actually
denies   that ``the probability of a $6$ is a physical property of a
given die \dots (including of course the total setup)''.   Such a
position would have some very peculiar consequences.

The pure frequentist position would (presumably) be that a
probability just is a (limiting) relative frequency in some 
ensemble:  absolutely
any ensemble, irrespective of how it is defined.  It could
be the ensemble which consists of all  the throws of a particular
coin.  But it could equally well be the ensemble which consists of
$10^4$ throws of one coin, followed by $10^4$ throws of a 
different coin.  It could even be the ensemble which consists of
$10^4$ throws of a coin, followed by $10^4$ Stern-Gerlach
measurements.

Suppose one really did make no distinction between these cases.
Then it would not only be  legitimate to argue
\begin{quote}  because
this coin has come up heads on approximately $50\%$ of the
last
$10^4$ tosses, therefore this same coin may be expected to
come up heads on approximately $50\%$ of the next
$10^4$ tosses.
\end{quote}
  It would be equally legitimate to argue 
\begin{quote}  because
this British penny  has come up heads on approximately $50\%$ of
the last
$10^4$ tosses, therefore that American dollar may be
expected to come up heads on approximately $50\%$ of the next
$10^4$ tosses.
\end{quote} It would even be legitimate to argue
\begin{quote}  because this British penny has come up heads on
approximately
$50\%$ of the last
$10^4$ tosses, therefore a Stern-Gerlach arrangement may be
expected to give the result ``spin up'' in approximately 
$50\%$ of the next
$10^4$ measurements.
\end{quote}
I believe that no frequentist would argue like this.  I think that
must mean that every supposed frequentist is in fact  tacitly
operating with some kind of propensity notion.

None of this detracts from the importance of Popper's shift to a
propensity approach.  Popper took an idea which, though tacitly
present all along, had never been sufficiently emphasized, and he
placed it centre stage.  This was a significant step.

The concept of a propensity is clearly implied by the way that
physicists talk. 
 For instance  half-lives are typically tabulated next to 
masses, as if they were just one more physical property.  However, I do
not think it is simply a matter of language.  It seems to me that the
concept   plays an essential role in the internal logic of current
physical theories. 

Suppose, for instance, Alice takes a large sample of
$\boldsymbol{\mathstrut^{228}\mathrm{Ac}}$ nuclei in her laboratory
in
\textbf{London}, and finds that approximately $50\%$ of them have
decayed after
$6$ hours.  Then she can legitimately infer that
\begin{quote}
  approximately $50\%$ of a sample of 
$\boldsymbol{\mathstrut^{228}\mathrm{Ac}}$ nuclei in
\textbf{Calcutta} may be expected to decay in $6$ hours.
\end{quote} But she cannot legitimately infer that 
\begin{quote}
  approximately $50\%$ of a sample of 
$\boldsymbol{\mathstrut^{226}\mathrm{Ra}}$ nuclei in
\textbf{London} may be expected to decay in $6$ hours.
\end{quote} 

The fact that this point is obvious should not be seen as detracting from
its importance.  Basic logical principles generally do seem obvious.
Suppose one wanted to programme a robot to understand logical arguments,
and perform physical experiments.  Then the robot would go badly wrong if
one failed to programme it with \emph{modus ponens} (the principle that, 
if
$P$ is true, and
$P\Rightarrow Q$, then $Q$ is true).  It would also go
badly wrong if one failed to programme it with the information that a
half-life is tied to the nuclear species, not to the place it was
measured.

Of course, in other contexts a probability can be tied to a
spatio-temporal location.  This happens in quantum field theory, for
example.  The point I am making is simply that the idea of a probability
being logically tied to
\emph{some} non-probabilistic physical entity plays an essential 
logical\footnote{
  The reader may question my use  of the word ``logical''.  
  Science depends on the procedure  by which we uncertainly infer, from
observations 
  of one set of events, predictions about other events. Since
it is a
 matter of drawing inferences  this procedure may fairly be
described as ``logical''.  

 Keynes~\cite{Keynes} attempted to explicate probabilistic reasoning
in terms of a novel kind of non-deductive logic.
 Ramsey~\cite{Ramsey} criticized that
idea.  In so far as it is directed at Keynes's \emph{specific}
proposal it seems to me that Ramsey's criticism is justified.  On the
other hand Ramsey (possibly) and de Finetti~\cite{Fin1} (definitely) have
suggested that probabilistic reasoning can be based \emph{purely}
on the principles of deductive logic (\emph{via} the Dutch book
construction).  I am not persuaded by that proposal either.

At the beginning of the $1930$'s
Wittgenstein~\cite{WittgensteinA,WittgensteinB} took to using the word
``grammar'' in preference to the word ``logic'' (in this connection it
may be worth noting that Wittgenstein cites Ramsey as a major influence
on his thinking in the preface to Wittgenstein~\cite{WittgensteinC}).  I
choose not to use the word ``grammar'' here because grammar has nothing
specially to do with reasoning, and because grammatical principles are,
to a considerable extent, arbitrary (\emph{pace} Chomsky). 
Nevertheless,  the concept I have in mind is at least as close to the
concept Wittgenstein intends by the word ``grammar'' as it is to the
concept Keynes intends by the word ``logic''.
 } 
role in our
current physical theories.  And it is that concept, of a 
logically
 bound
probability, which I take to be what is essentially meant by the term
``propensity''.

As I am presenting it here the concept of a propensity is, first and
foremost, a logical concept.   Under the influence of Popper  philosophers
have tended to view propensities as physically real properties.   But the
logical concept, as I have formulated it here, is consistent with an
epistemic point of view.

The trouble with thinking of propensities in the way that Popper suggests
is that it prompts questions like:  
what exactly is the difference between (1) a
$\mathstrut^{228}\mathrm{Ac}$  nucleus which decays after 
$26$ hours in spite of having had
a propensity to decay much sooner and (2) a
$\mathstrut^{228}\mathrm{Ac}$ nucleus which simply decays after that many
hours, without being troubled by any conflicting propensity? It seems
clear that any difference there may be is unobservable.

If one looks at propensities as objectively real properties they are
likely to seem empirically irrelevant.  If however, one looks at them
from a logical 
perspective\footnote{
  In contrasting this with the idea that a propensity is an objectively
  real property I do not mean to suggest that it is therefore
subjective.
  I think the subject-object dichotomy is potentially very misleading. 
I do not believe  the world genuinely is sundered, absolutely, quite
in the manner this terminology suggests.  Like Bohr I think that  is
the single most important lesson of quantum mechanics.

Wittgenstein~\cite{WittgensteinA} (pp.126--7) asks ``How do I know that
the colour red can't be cut into bits?'' This is surely not an empirical
observation. However, although the proposition can hardly be described as
an observed fact, or something inferred from observed facts, it seems to
me that it cannot appropriately be described as ``subjective''.

I would argue that logical  relationships are constitutive of
what we normally think of as reality (\emph{c.f.}, for example,
Wittgenstein~\cite{WittgensteinB}, p.116, where he speaks of the
``logical form'' of a patch in the visual field).
  }  
then it can be seen that they are highly relevant,
even though they are unobservable.

It is not, in general, necessary for an entity to be directly observable
in order for it to be empirically relevant.  It is enough that it be
non-redundantly embedded in a structure of empirical thought.  The
state vector is not directly observable; nor are logical relations.  But
they could not be considered empirically irrelevant.

\section{Retrodictive Inferences:  the Bayesian Methodology}
\label{sec:Bayes}
I argued in Section~\ref{sec:frequentism1}  that retrodictive
inferences depend on background assumptions.  Those assumptions are
themselves probabilistic in character.  Their role emerges most
clearly in the Bayesian approach, discussed in  this
section.  I will discuss the orthodox theory of statistical inference in
the next section.

The term ``Bayesian'' tends to be associated with an epistemic
interpretation of probability.  However, there are  objectivists who
favour the Bayesian methodology.  I argued  that von Mises, though usually
described as a frequentist, is really a propensity theorist.  He is,
besides, a  Bayesian (see von Mises~\cite{Mises} pp.~117--25, 157--9 and
von Mises~\cite{MisesB}, chapters VII and X).

I myself take an epistemic view. 
However, for the purposes of this paper it is more appropriate to present
the Bayesian methodology from an objectivist perspective, such as that of
von Mises.  I want to expose the deficiencies of the objectivist idea.  
The best way to do that is to adopt  objectivist assumptions, and  see
where they lead.

Also, I want to compare the Bayesian approach with what is now the
orthodox approach.  This was developed by Fisher and others in the first
few decades of the last century in conscious opposition to the Bayesian
methodology.  If one wants to understand Fisher's reasons for rejecting
the Bayesian approach one needs to look at it through objectivist eyes.

Let
$s_i=0$ (respectively
$1$) if the coin comes up tails (respectively heads) on the
$i^{\mathrm{th}}$ toss.   Let  $\mathbf{s}=(s_1,s_2,\dots, s_n)$  be the
sequence describing the result of the first $n$ tosses.  Let $p$ be the
true probability of heads (I am following von Mises's objectivist account,
so I am assuming that there does exist a real, objective probability of
heads).  Then, if the tosses are independent, the  
probability of obtaining the sequence $\mathbf{s}$ is
\begin{equation}
P(\mathbf{s}|p) = p^{h(\mathbf{s})} (1-p)^{n-h(\mathbf{s})}  
\label{eq:Bayes1}
\end{equation}
where $h(\mathbf{s})=\sum_{i=1}^{n} s_i$ is the number of heads.

Now if it was a certain fact that the coin was fair  we would have 
$p=1/2$ and every sequence $\mathbf{s}$ would have the same probability $
1/2^n$.    Suppose, however,  that the coin  was randomly selected
from a large population of coins, not all of which are fair.  Let
$f_{\mathrm{i}}(p) dp$  be the probability that  the probability of heads
is in the interval
$(p,p+dp)$.  Then the unconditional probability of obtaining the sequence
$\mathbf{s}$ is 
\begin{equation}
P(\mathbf{s})  = \int_{0}^{1} P(\mathbf{s}|p) f_{\mathrm{i}}(p) dp\,.
\label{eq:Bayes2}
\end{equation}
Suppose, now, that we do toss the coin $n$ times, and suppose we obtain
the sequence
$\mathbf{s}$.  Then, by an application of Bayes's theorem,  the  
probability that the probability of heads is in the  interval
$(p,p+dp)$, conditional on this new information, is
\begin{equation}
P(p|\mathbf{s})dp = \frac{P(\mathbf{s}|p) f_{\mathrm{i}}(p) dp}{
\int_{0}^{1} P(\mathbf{s}|p) f_{\mathrm{i}}(p) dp}\, .
\end{equation}
Setting $f_{\mathrm{f}}(p)=P(p|\mathbf{s}) $ this gives, in view of 
Eq.~(\ref{eq:Bayes1}),
\begin{equation}
f_{\mathrm{f}}(p) = K p^{h(\mathbf{s})} (1-p)^{n-h(\mathbf{s})}
f_{\mathrm{i}} (p)
\label{eq:Bayes3}
\end{equation}
where $K$ is a normalization constant.
$f_{\mathrm{i}} (p)$ is usually called the prior (or
\emph{a priori})
distribution, and $f_{\mathrm{f}} (p)$ the posterior (or
\emph{a posteriori}) distribution.

It will be observed that von Mises explicitly appeals to the concept of a
second-order probability:  a probability of a probability.
Savage~\cite{Savage} (p.~58), among others, has criticized this idea. 
However, if one accepts  von Mises's objectivist viewpoint the concept
must be regarded as 
legitimate\footnote{
Of course,  from de Finetti's~\cite{Fin1}  viewpoint
  it makes no sense to interpret $f_{\mathrm{i}}(p)dp$ as a probability
  (on de Finetti's
  assumptions a second order probability would have to represent a belief
  about  one's own belief, not a belief about the object of interest).  de
  Finetti has found a most ingenious way round this difficulty.  In 
  his scheme the weight function 
  $f_{\mathrm{i}}(p)$ 
   \emph{encodes} a probability distribution, without itself \emph{being}
  a probability distribution.
 }.  
As we saw von Mises, though usually
described as a frequentist, is in fact a propensity theorist.  That is he
thinks that the probability of a  coin coming up heads is a physical
property of that particular coin (plus the tossing method) (von
Mises~\cite{Mises}, p.~14).  If one makes that assumption, and if it is
legitimate to consider the probability that the mass $m$ of a randomly
selected coin lies in a certain interval, then it must be equally
legitimate to consider the probability that the probability $p$ of the
coin coming up heads lies in  a certain  interval. 

To understand the significance of Eq.~(\ref{eq:Bayes3}) (as von Mises
sees it) consider, for simplicity, a case where there are only two
possible values of
$p$.   Imagine a bag containing a large number of fair coins,
with $p=0.5$, and a much smaller number of biased coins, with $p=0.9$. 
For the sake of definiteness suppose that the proportion of biased coins
is
$10^{-4}$.

Now consider the experiment which consists in shaking the bag, selecting
a coin at random, tossing the coin $30$ times, and then replacing it.  
Suppose this experiment is repeated infinitely many times.  Let
$S_{\mathrm{i}}$ be the  set 
 of all such experiments, and let 
$S_{\mathrm{f}}$ be the subset obtained by selecting just those
cases where the coin came up heads on each of the $30$ tosses.
Then $S_{\mathrm{i}}$ is described by the distribution
\begin{equation}
f_{\mathrm{i}} (p) = 0.9999\, \delta(p-0.5)+ 0.0001\, \delta(p-0.9)
\label{eq:CoinsInABagA}
\end{equation}
while  $S_{\mathrm{f}}$ is described by the distribution
\begin{equation}
f_{\mathrm{f}} (p) = 0.0002\, \delta(p-0.5)+ 0.9998\, \delta(p-0.9)
\end{equation}
(as follows from Eq.~(\ref{eq:Bayes3})).  In the set of all experiments
the coin is fair in $99.99\%$ of cases. But in the subset of experiments
in which  the coin comes up heads on each of
$30$ tosses, the coin is biased in $99.98\%$ of cases.

It will seen from Eq.~(\ref{eq:Bayes3}) that, in a Bayesian inference,
the conclusion to the argument (the distribution
$f_{\mathrm{f}}$) is produced by the interplay between a  probabilistic
premise (the distribution
$f_{\mathrm{i}}$) and a set of factual observations (the sequence
$\mathbf{s}$).  It is this interplay which explains the point I made in
Section~\ref{sec:frequentism2}, that some observations force a major
change in our beliefs while others, though equally improbable on our
starting assumptions, do not.

It seems to von Mises that the Bayesian methodology provides a
perfectly clear, fully objective theory of retrodictive probabilistic
reasoning.  He is at a loss to understand why it is not more widely
accepted~\cite{Mises} (pp.158--9):
 \begin{quote}
I do not understand the many beautiful words used
by Fisher and his followers in support of the likelihood theory \dots

\medskip

\noindent
\dots  We can only hope that statisticians will return to
the use of the simple, lucid reasoning of Bayes's conceptions.
\end{quote}
de Finetti's  epistemic interpretation of probability is at the opposite
pole from von Mises's objectivism.  However, he and von Mises are at one
in their attitude to orthodox statistics.  de Finetti~\cite{Fin1} (p.~245)
says, for example,
\begin{quote}
``Those who reject the Bayesian approach cannot base their inferences on
the posterior distribution even if they wished to---it does not make sense
so far as they are concerned.  As a result, they are forced to have
recourse to \emph{ad hoc} criteria, and hence to open the floodgates to
arbitrariness. \dots The best they can \dots do is to base themselves on
the likelihood function; failing that, they simply resort to playing with
formulae that are without any real foundation.''
\end{quote}
I will argue below that these criticisms of the orthodox methodology are
 justified.  However, I first want to
examine Fisher's reasons for rejecting the Bayesian approach.

Fisher~\cite{FisherStat} (also see Fisher~\cite{FisherDesign}) gives a
 detailed discussion of  the Bayesian 
methodology\footnote{ 
  von Mises~\cite{Mises} (p.~159) says ``Fisher emphatically avoids all
reference to Bayes's solution of the problem of inference; this is for
him a matter of principle''. von Mises presumably wrote this for the
first edition of his book, before he had read Fisher~\cite{FisherStat}. 
At any rate, it is a misapprehension. Fisher does not dismiss the
Bayesian methodology out of hand, without argument.
  }.   He 
is at pains to defend it against most of the attacks which have been made
on  it.  He argues~\cite{FisherStat} (pp.~8--38) that criticism
has mainly been directed at  inappropriate applications of the method,
which fail to respect the conditions of Bayes's theorem, and that the
method itself is perfectly sound. He also  says~\cite{FisherStat} (p.~48)
that there can be no fundamental objection to second-order
probabilities:  probabilities of probabilities. His only objection to the
Bayesian method is that the  prior
$f_{\mathrm{i}}$ is, in practice, usually unknown.  This means that, in
most cases, Eq.~(\ref{eq:Bayes3}) merely expresses one unknown in
terms of another.

It is generally true that the Bayesian can only get a
probabilistic judgment \emph{out}, as the conclusion to a piece of
retrodictive reasoning, if s/he begins by feeding a probabilistic
judgment \emph{in}, as an initial assumption.  Fisher's point is that, in
most (though not all---see Fisher~\cite{FisherStat}, pp.~18--20 and
127--132) applications, that initial assumption cannot be based on known
facts about the objective situation.   In his view that makes the
Bayesian method, in most applications, scientifically useless.

von Mises~\cite{MisesB} (pp.~339--45) (also see von Mises~\cite{Mises},
pp.~122--4) has a response to this objection.
 If $n$ is large then
$p^{h(\mathbf{s})} (1-p)^{n-h(\mathbf{s})} \approx K' \delta (p-p_0)$ 
where
$p_0=h(\mathbf{s})/n$ and $K'$ is a normalization constant.  Inserting
this expression in Eq.~(\ref{eq:Bayes3}) one obtains, on the assumption
that  $f_{\mathrm{i}}$ satisfies certain conditions (which von Mises
explicitly states)
\begin{equation}
f_{\mathrm{f}}(p) \approx \delta(p-p_0)
\label{eq:von MisesLargeNumbers}
\end{equation}
independently
 of $f_{\mathrm{i}}$.  von
Mises infers that if one obtains $r$ heads in $n$
coin tosses, and if $n$ is sufficiently large, then it is nearly certain
that
$p\approx r/n$.  

His response is inadequate for two reasons.  In the first place
$f_{\mathrm{i}}$ might not satisfy von Mises's conditions.  If, for
example, $f_{\mathrm{i}} (p) = \delta(p-0.5)$, then $f_{\mathrm{f}} (p) =
\delta(p-0.5)$ whatever the values of $r$ and $n$. von Mises would
doubtless argue that his conditions are very plausible.  However,  the
fact is that they are empirically unfounded.  No set of observations
could ever contradict the assignment $f_{\mathrm{i}} (p) = \delta(p-0.5)$.

The second problem is that, even if we accept von Mises's conditions,
there is no empirical  criterion to tell us how large $n$ must be for it
to be nearly certain that $p\approx r/n$.  Suppose, for instance, we have
obtained $60$ heads in $100$ tosses.  For some choices of
$f_{\mathrm{i}}$ this would be very strong evidence that the coin is
biased; for others it would not.  The question, as to which alternative
applies, cannot be decided empirically.

In the case of a die von Mises~\cite{Mises} (p.~123) suggests that $500$
throws should be enough to give a reliable estimate of the true
probabilities. But he does not adequately explain where this
number is coming from.  It cannot be inferred just from the observations,
without additional assumptions.

Fisher, like von Mises, takes an objectivist view. I think he must be
correct to think that,  seen from that perspective, this feature of the
Bayesian methodology is unacceptable.  The trouble is that the orthodox
methodology, which Fisher and others devised in an attempt to get round
the problem, does no better.

\section{Retrodictive Inferences:  the Orthodox Methodology}
\label{sec:orthodox}
A Bayesian argument takes an initial probability assignment $P$ (the
 prior $f_{\mathrm{i}}$), adds to it some factual data $F$, and
from this derives a new probability assignment
$Q$ (the posterior $f_{\mathrm{f}}$). Formally:
\begin{equation}
P\wedge F \Rightarrow Q\, .
\label{eq:BayesGeneralForm}
\end{equation}
Fisher's problem is  that all we  observe is the  factual data $F$.  He
consequently thinks that, if probability assignments are not to 
float free of any empirical attachment, it is essential that Bayesian
inferences of the form~(\ref{eq:BayesGeneralForm}) be supplemented with a
new kind of inference having the general form
\begin{equation}
F\Rightarrow Q
\label{eq:OrthodoxHopedForm}
\end{equation}  
where a probability 
assignment
$Q$ is inferred directly from the factual
data $F$, without the assistance of any prior probabilistic 
assumption (see, for example,  Fisher~\cite{FisherStat}, pp.~54--5).
In this section I hope to convince the reader that valid inferences of
this form do not exist.

I believe that Fisher is entirely correct in thinking that the
dependence of Bayesian inferences on prior probabilistic assumptions is a
serious difficulty for one of his philosophical persuasion.  Where he
goes wrong is in thinking that this is a reason for rejecting the Bayesian
methodology.  It is not just Bayesian inferences which have the 
form of Eq.~(\ref{eq:BayesGeneralForm}), but
\emph{every}  retrodictive probabilistic inference, without exception.

Consider
\begin{quote}
\textbf{Alice's argument:}  Alice spins a roulette wheel once and obtains
the number $13$.  She concludes that the wheel is
fair.
\end{quote}
This argument is clearly invalid.  A single spin of a roulette wheel
tells one virtually nothing about the underlying probability
distribution.  The fact that the number $13$ came up once, in a single
spin of the wheel, does not imply that the other numbers are even
possible, much less that they each have probability $1/37$.  It also
seems a little strange to argue that, because $13$ did occur, therefore
$13$ is not very likely to occur---an anti-inductive argument, as it
might be called.

Now compare
\begin{quote}
\textbf{Bob's argument:}  Bob tosses a coin $1000$ times and obtains
$500$ heads.  He concludes that $(0.459,\, 0.541)$ is a $99\%$ confidence
interval for the probability $p$ of the coin coming up heads.
\end{quote}
It may appear that Bob has solid reasons for this conclusion. But in fact,
if Bob is claiming to extract his conclusion just from the empirical
data, without additional assumptions, then his claim is no better founded
than Alice's.

A sequence of $1000$ coin tosses is equivalent to $1$ spin of a big
roulette wheel, divided into $2^{1000}$  sectors.  Let $\mathbf{b}$ be the
particular sequence which Bob obtains.  Then, on the basis of one spin of
the equivalent roulette wheel, Bob is arguing
\begin{quote}
 Because sector $\mathbf{b}$ \emph{did} occur, therefore  each of the
 sectors which \emph{did not} occur
 has probability $\ge 6\times 10^{-339}$
\end{quote}
He is also arguing anti-inductively:
\begin{quote}
  Because $\mathbf{b}$ did occur, therefore $\mathbf{b}$ is very
unlikely to occur.
\end{quote}
If Bob really was basing himself just on the observed facts, and
nothing else whatever, his argument would have the same extraordinary
features  as Alice's argument.   It would clearly be invalid.

Of course, Bob is not really basing himself just on the factual data.
He is supplementing the
factual observation
\begin{quote}
$F=$ ``coin  tossed $1000$ times and  sequence $\mathbf{s}$
obtained''
\end{quote} 
with the prior probabilistic assumption
\begin{quote}
$P=$ ``tosses are independent, and  probability of heads is
constant''
\end{quote}
$P$ is logically equivalent to the statement
that, for some fixed $0\le p \le 1$, the probability of obtaining
an arbitrary sequence $\mathbf{s}$ is
\begin{equation} 
 P(\mathbf{s}) 
= p^{h(\mathbf{s})} (1-p)^{1000-h(\mathbf{s})}
\label{eq:TossDistributionOrthodox}
\end{equation} 
(where
$h(\mathbf{s})$ is the number of heads in the sequence
$\mathbf{s}$, as before).

The set of all probability distributions on the space of 
sequences of $1000$ heads and tails is a $(2^{1000}-1)$ parameter family.
The assumption $P$ restricts the class of admissible distributions to the
one parameter family specified by
Eq.~(\ref{eq:TossDistributionOrthodox}).  That is a  severe
restriction.  Without such  a restriction no valid,  non-trivial
retrodictive inference is possible.

If the class of admissible distributions is suitably restricted, then one
can  legitimately  draw  probabilistic conclusions from a
single spin of an ordinary roulette wheel.

To see this  let $H_l$ be the
hypothesis
\begin{equation}
P(r)=\begin{cases}
 \frac{1}{10} \qquad & \text{if $l\le r \le  l+9$} \\
  0 \qquad  &\text{otherwise}
\end{cases}
\end{equation}
where $l=0,1,\dots, 28$ and $P(r)$ is the probability of obtaining the
number $r$ when the wheel is spun once.

 Now let
\begin{quote}
$F=$``wheel was spun once and  number $13$ was obtained''
\end{quote} 
and let $P$ be the  disjunction
\begin{quote}
$P= \bigvee_{l=0}^{28} H_l$
\end{quote}
$F$ on its own has virtually no
probabilistic implications (beyond the obvious implication, that $13$ was
possible and that none of the other numbers was certain).
However, the conjunction $P\wedge F$ 
does
have substantial probabilistic
implications\footnote{
  \label{fn:zeroProb}
  At least, it does if one is allowed to assume
 that  zero probability
  events are impossible.
}:  
namely, the proposition $Q= \bigvee_{l=4}^{13} H_l$.

If Alice is permitted to assume $P$ then she can validly argue, for
example,
\begin{enumerate}
\item[(a)] because $13$ did occur, therefore   $12$ and $14$ are not both
impossible.
\end{enumerate}
and
\begin{enumerate}
\item[(b)] because $13$ did occur, therefore $13$ has probability $1/10$.
\end{enumerate}
However, she is only getting these probabilistic conclusions
out  because she began by feeding a probabilistic
assumption in.  In particular, the superficially
anti-inductive character of inference (b) is not mysterious. 
Alice only arrives at the conclusion, that $13$ is  unlikely to
occur,  because she began by assuming that $13$ is  unlikely to
occur.  If $P$ is true, then the probability of $13$ is necessarily 
$\le 1/10$.  Alice is, in fact, assigning the maximum probability
consistent with her  starting assumptions.

This example---Alice's modified argument, as I will  call it---is
admittedly artificial.  However, the same idea of assuming
a  disjunction $P$, and then using the observations to narrow it down
to a smaller disjunction $Q$ is at the root of  many, if not all
orthodox statistical inferences.

Let us go back to the example of $1000$ coin tosses.  Let $H_p$ be the
hypothesis ``tosses are independent, and the probability of heads is $p$
on every toss''.  If
$H_p$ is true the probability of obtaining the sequence $\mathbf{s}$ with
$h(\mathbf{s})$ heads is $p^{h(\mathbf{s})} (1-p)^{1000-h(\mathbf{s})}$. 
Unlike the distributions in Alice's modified argument, this distribution
is everywhere non-zero (unless $p=0$ or
$1$).  However,   it is  sharply peaked at $h(\mathbf{s})=1000 p$. 
Orthodox statisticians take it that, away from this peak, the distribution
may be regarded as \emph{effectively} zero.  They think this entitles
them to proceed in the same manner as Alice, in her modified argument. 
They begin by assuming the disjunction $P=\bigvee_{p\in [0,1]} H_p$, and
 then use the observed sequence $\mathbf{s}$ to narrow this down to a
disjunction  $Q=\bigvee_{p\in I_c} H_p$, where $I_c$ is a confidence
interval.  

For example, suppose as before that a sequence $\mathbf{b}$ containing
$500$ heads is obtained.   Let
$E_{-}$ (respectively
$E_{+}$) be the event that  the  number of heads is $\le
500$ (respectively  $\ge 500$).  Let $P(E_{\pm}|H_p)$ be  the conditional
probability of
$E_{\pm}$ given $H_p$.  Then 
\begin{equation}
 P(E_{+}|H_p) \le 0.005
\label{eq:ConfIntLow}
\end{equation}
when $p\le 0.459$ while
\begin{equation}
 P(E_{-}|H_p) \le 0.005
\label{eq:ConfIntUpp}
\end{equation}
when $p\ge 0.541$.  If $P(E_{+}|H_p) =0$ whenever $p\le 0.459$ and 
$P(E_{-}|H_p)=0$ whenever $p\ge 0.541$ we
would be in the same situation as Alice, in her modified argument.  We
could conclude that 
$p$ \emph{certainly} $\in (0.459, \, 0.541)$ (\emph{modulo} the
qualification in footnote~~\ref{fn:zeroProb}).  As it is the conditional
probabilities $\neq 0$, and so certainty would not be justified. 
The probabilities are, nevertheless, small.   Orthodox  statisticians
consider that this entitles us to be, in some sense,  confident that
 $p\in (0.459, \, 0.541)$.

The orthodox analysis depends, not just on the factual data, but also on
the probabilistic assumption $P=\bigvee_{p\in [0,1]} H_p$.  This
assumption is no more empirical than the choice of Bayesian prior.

In Section~\ref{sec:frequentism2} I appealed to the idea that a coin might
be systematically biased in favour of heils:  \emph{i.e.}\ biased
towards heads on some tosses and tails on others.  This suggestion may
have seemed fanciful.  But it does in fact describe what a classical
determinist like Laplace~\cite{Laplace} believed to be true in objective
reality.  Laplace thought that for a superhuman being, who had complete
knowledge of the objective situation, the probability of heads on any
particular toss  would always be either $0$ or $1$.  In other words, he
thought that if ``heils'' and ``taads'' are
appropriately defined then, for such a being, the  probability of heils is
always
$1$.

The de Broglie-Bohm theory~\cite{Bohm,Holland} indicates  that the
empirical predictions of quantum mechanics are consistent with complete
determinism at the micro-level.
 Consider, for example, a measurement of
$\hat{\sigma}_z$ on a succession of particles each initially in an
eigenstate of
$\hat{\sigma}_x$.  We cannot, at present, empirically exclude the
possibility that, for each particle that passes through the apparatus,
the measurement outcome is fully determined by the initial conditions. 
Consequently there is, at present,   no
way to empirically decide between the hypotheses:
\begin{enumerate}
\item[$A$:]  For every particle the probability of obtaining the result
``spin-up'' is
$1/2$
\item[$B$:]  For every particle the probability of obtaining the result
``spin-up'' is either $0$ or $1$.
\end{enumerate}
Nor is there any obvious way to empirically exclude
\begin{enumerate}
\item[$C$:]   For every particle the probability of obtaining the result
``spin-up'' is either $0.4$ or $0.6$.
\end{enumerate}
---not to mention other, more complicated possibilities.

Analogous considerations apply to the proposal that the  measurement
outcomes are statistically independent.  Let $X_n$ be the
$n^{\mathrm{th}}$ measurement outcome.  There is no obvious way  to
empirically discriminate
\begin{enumerate}
\item[$A'$:]  $P(X_{2n}=X_{2n+1})=1/2$ for all $n$.
\item[$B'$:]  $P(X_{2n}=X_{2n+1})=0$ for some values of $n$ and $1$ for
others
\item[$C'$:]  $P(X_{2n}=X_{2n+1})=0.4$ for some values of $n$ and $0.6$
for others.
\end{enumerate}
(not to mention other, more complicated possibilities).

Similar remarks apply to the coin-tossing example. 

It may be suggested that the assumption $P=\bigvee_{p\in [0,1]} H_p$
seems very plausible.  I would agree with that: it seems very plausible
to me also.  However, that has no bearing on my argument here.  I am here
only making the simple logical point, that $P$ represents an additional
assumption, not contained in the empirical data.

Hume~\cite{HumeA,HumeB}  argued that we have no sufficient 
empirical reason for expecting the sun to  rise tomorrow.  The contrary
proposal, that the sun  will probably not rise tomorrow, does---of
course---seem very implausible.  But Hume does not deny that.  His point
is not that our ordinary belief is not very plausible, but only that it
cannot logically be derived purely from the observed facts, without any
non-empirical input.  The same is true of  conclusions reached by the
kind of generalized inductive argument considered  here.

One might try to argue that the assumption $\bigvee_{p\in [0,1]} H_p$
can be justified by appealing to the results of \emph{previous}
coin-tossing experiments.  However, any conclusions drawn from those
previous experiments would  themselves  have to depend on previous prior
assumptions.  Trying to  justify probability by probability is like
trying to justify induction by induction:  it cannot be done, unless the
pump has first been primed, by some initial non-empirical assumption.

\section{Orthodox \emph{v.} Bayesian}
\label{sec:orthodoxVbayes}

In the last section I argued that the orthodox methodology relies on
prior, non-empirical assumptions, just like the Bayesian methodology. 
However, it may  look as though the orthodox methodology is still 
preferable  because it does not require us to make
\emph{so many} such assumptions.
As we saw in Section~\ref{sec:Bayes}, von Mises's objectivist 
version of
the Bayesian 
methodology\footnote{
   In de Finetti's~\cite{Fin1} epistemic
   version Eq.~(\ref{eq:Bayes3}) is derived by a different route, 
 in which $\bigvee_{p\in [0,1]} H_p$ is replaced by the assumption
  of exchangeability.  The end mathematical result is the same, but 
  the conceptualisation is quite different 
    (see, however, Howson
   and Urbach~\cite{Howson},  pp.~232--3).
}, 
 as applied to the coin-tossing example, relies on the prior assumption
$\bigvee_{p\in [0,1]} H_p$, just as the orthodox approach does.  But it
also relies on the prior distribution $f_{\mathrm{i}}$.  By
contrast, it may appear  that the orthodox approach does not make any
non-empirical assumption additional to $\bigvee_{p\in [0,1]} H_p$.

However, it will be found on closer examination that this is not correct.
Assumptions corresponding to the  distribution $f_{\mathrm{i}}$
play an essential role in the orthodox approach.  The only difference is
that, whereas in the Bayesian approach these assumptions are explicitly
built into the formalism, in the orthodox approach they are tacit, and
often unrecognized. All that the orthodox statistician achieves by
suppressing the function 
$f_{\mathrm{i}}$ is to produce a misleading appearance of greater
objectivity, at the price of a serious loss of logical coherence.  In
particular, the orthodox methodology obscures the point which emerged
from the discussion in  Section~\ref{sec:frequentism2}: that retrodictive
inferences are critically dependent on our background knowledge and
beliefs.

I noted at the end of Section~\ref{sec:Bayes} that the Bayesian approach
fails to give a purely objective criterion for deciding how many tosses
are needed to tell whether a coin is biased.  Suppose, for instance, a
coin comes up heads on $60$ out of $100$ successive tosses. For some
choices of prior distribution $f_{\mathrm{i}}$ this will imply that there
 probably is a substantial bias, for others it will not.  On these grounds
Fisher rejects the Bayesian methodology.  Yet the alternative methodology
which he advocates is no more objective.

On the hypothesis that the coin is fair (and the tosses independent) the
probability  that heads will come up more than $60$ times in $100$
successive tosses is  $0.028$.  So if we perform a one-tailed test the
hypothesis, that the coin is fair, will be rejected at the $95\%$ level,
but not at the
$99\%$ level of significance. If, on the other hand, we perform a
two-tailed test the hypothesis will not  be rejected even at the $95\%$
level (though it will be rejected at the $90\%$ level).  So do we accept
that the coin is biased or not?  That, it seems, is up to the subjective
decision of the statistician.

Fisher~\cite{FisherStat} (p.~45) has this to say, regarding the choice of
significance level:
\begin{quote}
``no scientific worker has a fixed level of significance at which from
year to year, and in all circumstances, he rejects hypotheses; he rather
gives his mind to each particular case in the light of his evidence and
his ideas''
\end{quote}
In other words, the choice of significance level depends on exactly the
same factors which determine the choice of Bayesian prior:  namely, the
background knowledge and beliefs of the statistician.  Similarly with the
decision as to whether to use a one-tailed or a two-tailed test.

The orthodox methodology might be considered superior to the Bayesian
methodology if the statement, that $H$ is rejected at the $95\%$
level, meant that $H$ is false with probability $\ge 0.95$.  However,
the statement cannot validly be interpreted in this way,
as orthodox statisticians are at pains to emphasize.  A significance level
is not a probability (as is already apparent from the fact that it
depends on whether the test is one-tailed or two-tailed).

The fact that orthodox statistical conclusions are not purely objective
tends to be obscured by the fact that orthodox statisticians standardly
choose to work at only a small number of different signficance levels
(typically $95\%$ or $99\%$).  However, the fact that everyone judges the
same is not, in itself, a reason for taking a judgment to be
objective.  In any case, Bayesian statisticians could achieve an
equal degree of unanimity by the simple expedient of always working in
terms of the uniform prior
$f_{\mathrm{i}}(p)=1$  (as, indeed, was originally recommended
by Bayes and Laplace~\cite{Laplace}).

The orthodox methodology is not superior in point of objectivity.  On the
other hand it is clearly inferior in point of logical cogency.

The argument we have been considering may be summarized as follows: 
\begin{enumerate}
\item[] \medskip \textbf{Argument 1:} \medskip
\item[] If the coin is fair (and the tosses  independent) the probability
of more than $60$ heads in $100$ tosses is $\le 0.05$. 
\smallskip
\item[] The coin did came up heads on $60$ out of $100$ tosses.
\medskip
\item[] \emph{therefore}
\medskip
\item[] The hypothesis, that the coin is fair, is rejected at
the $95\%$ level.
\end{enumerate}
Now campare this with the example I discussed in
Section~\ref{sec:frequentism2}, where Alice wins a lottery
having $2^{24}$ tickets.  Suppose one were to reason as follows:
\begin{enumerate}
\item[] \medskip \textbf{Argument 2:} \medskip
\item[] If the lottery is fair the probability of Alice winning 
$=6\times 10^{-8}$.
\smallskip
\item[] Alice did  win.
\medskip
\item[] \emph{therefore}
\medskip
\item[] The hypothesis, that the lottery is fair, is  rejected at the
$99.99999\%$  level.
\end{enumerate}
If argument~$1$ is valid just as it stands (if the conclusion does not
tacitly depend on some additional, inexplicit asssumptions), then it is
hard to see what objection there can be to argument~$2$.

It might be suggested that argument $2$ is invalid because the conclusion
is based on a singular event.  However, as we saw in
Section~\ref{sec:frequentism2}, it is easy to think of cases where one
can validly draw retrodictive conclusions from singular occurrences. 
Besides, if one wants to erect it as an absolute principle, that
retrodictive conclusions must be based on repeated trials, one has to
decide just how many repetitions are needed.  It is hard to see how the
decision can be other than arbitrary.

In any case one can  find sufficiently many logical obscurities in
the orthodox analysis of the coin-tossing problem.  Argument~$1$ depends
on grouping the particular sequence containing $60$ heads which \emph{did}
occur together with all the  other sequences containing $60$ or more
heads which
\emph{did not} occur.  It is hard to see, on orthodox principles,
any compelling reason for adopting this procedure.  However, if we did try
basing ourselves just on the sequence which actually occurs it would lead
to some  strange conclusions.

Suppose, for instance, that a coin is tossed $100$ times, and a sequence
$\mathbf{s}$ containing $50$ heads is obtained.  This would usually be
seen as  favouring the hypothesis that the coin is fair. 
However, if retrodictive inferences really were based  just on ``the
resistance felt by the normal mind to accepting a story intrinsically too
improbable'' (Fisher~\cite{FisherStat}, p.~43) it is hard to see what
objection there could be to
\begin{enumerate}
\item[] \medskip \textbf{Argument 3:} \medskip
\item[] If the coin is fair (and the tosses independent) the probability
that  $\mathbf{s}$ will occur  $=7.9\times 10^{-31}$.
\smallskip
\item[]  $\mathbf{s}$ did occur.
\medskip
\item[] \emph{therefore}
\medskip
\item[] The hypothesis, that the coin is fair, is  rejected at the
$99.99 \cdots 9\%$  level.
\end{enumerate}
(see Howson and Urbach~\cite{Howson}, p.~123).

Of course, no orthodox statistician would argue in that fashion.  The
question  is:  why?  I do not see how it is possible, on orthodox
principles alone, to explain, in clear, logically compelling terms,
 why argument~1  is valid while arguments~2 and~3 are not.

Although Jaynes~\cite{JaynesB} has argued  that the
orthodox methodology is often not the most effective way to extract 
conclusions from statistical data, I believe no one has questioned the
actual validity of conclusions reached by orthodox means.  However it is,
I think, difficult to disagree with de Finetti~\cite{Fin1} (p.~245) when
he says that orthodox statistics relies on a multiplicity of \emph{ad
hoc}  decisions, whose logical basis is often far
from  clear\footnote{
  Howson and Urbach~\cite{Howson} (p.~124) say that Fisher is not really
  a falsificationist like Popper (as Gillies~\cite{GilliesA} (p.~147)
maintains), but rather a quasi-falsificationist.  I think that 
is probably a fair description, provided
``quasi-falsificationism'' is not taken to be a coherent system of
logical thought.
 }.

Jeffreys~\cite{JeffreysB} (p.~393), contrasting his Bayesian approach
with Fisher's orthodox one, comments 
\begin{quote}
  I have in fact been struck repeatedly in my own work, after being 
  led on general principles to the solution of a problem, to find that
  Fisher had already grasped the essentials by some brilliant piece of
  common sense.
\end{quote} 
In other words Fisher, notwithstanding
his lack of logical system,  generally gets the right answer due to the
power of his intuition (Jaynes~\cite{JaynesB} (p.~199) also pays
tribute to the depth of Fisher's intuitive insight).  
Bell~\cite{Bell} (p.~174) remarks that the Copenhagen formulation of
quantum mechanics,  in spite of the obscurity of its basic concepts, is
still enormously successful on a practical level due to the ``discretion
and good taste'' of its practitioners.  The orthodox approach to
statistics is equally  reliant on these  qualities of discretion and good
taste.  

If one was simply interested in practical problems of error-analysis,
epidemiology and the like, the orthodox approach might be
 satisfactory (see, however, Jaynes~\cite{JaynesB}).  However, the reader
of this article is likely to be interested in the foundational problems
of science (as, it should be said, is Fisher).   From that point of view
the logically unsystematic character of the orthodox approach is a
serious disadvantage, for it tends to obscure the real character of
probabilistic reasoning.  The Bayesian methodology is greatly preferable.

If one looks at it in Bayesian terms it is easy to see why the conclusion
to argument~1 is valid, whereas the conclusions to arguments~2 and~3 are
both invalid.

Let us first formulate the Bayesian approach in general terms.
Suppose we have a set of hypotheses $H_1,H_2,\dots,$ (for
simplicity assumed discrete), and suppose we assume that
$\bigvee_{i} H_i$ is true, so the prior probabilities satisfy 
$\sum_{i} P(H_i)=1$.  Let $E$ be the observed outcome. Then the
posterior  probability of $H_i$ given the
data
$E$ is
\begin{equation}
P(H_i|E) = \frac{P(E|H_i) P(H_i)}{P(E)}
\label{eq:bayes}
\end{equation}
where $P(E)=\sum_{i} P(E|H_i) P(H_i)$.  The  conditional
probability $P(E|H_i)$ is often called the \emph{likelihood} of 
$H_i$ relative
to the data $E$.  

There are two points to notice about this formula:  (1)
the posterior  depends on  an interaction between the
likelihoods
$P(E|H_i)$ and the priors $P(H_i)$; and (2) the factor
$P(E)$ in the denominator may be, and often is very small---which means
that the posterior $P(H_i|E)$ may be, and often is appreciable even when
the likelihood $P(E|H_i)$ is very small.  

Orthodox statisticians 
neglect both  these points.  Their  desire to  fit the
theory onto a Procustean bed of pure objectivism makes them  try to get
everything from the likelihoods $P(E|H_i)$ alone.  The effect is to
mutilate the logical structure of the theory.

Let us now specialize the formula to the case of Alice's lottery ticket. 
This will give a formal basis to the intuitive considerations of
Section~\ref{sec:frequentism2}.

 Let
$N$ be the number of tickets.  Let
$H_0$ be the hypothesis that the lottery is fair, and let $H_i$  be the
hypothesis that ticket $i$ is certain to win, for $i=1,2,\dots ,N$. 
Suppose we take $P(H_0) = 1 -\epsilon$ and $P(H_i) = \epsilon/N$ for $i\ge
1$ and some small $\epsilon$ (corresponding to a situation where we
think there is a small probablity of the lottery being rigged in
\emph{someone}'s favour, but have no idea who that someone might be).  
Let
$a$ be Alice's ticket, and let $E$ be the event that Alice's ticket
wins.  Then 
\begin{equation}
 P(H_0| E) = P(H_0) = (1-\epsilon)\, .
\end{equation} 
Under these conditions the event, that Alice wins the lottery, does not
change our assessment of the probability of the lottery being rigged---in
 agreement with ordinary intuition.  Note that $P(H_0|E)$ is close to $1$
even though the likelihood $P(E|H_0)$ is very small.  This is because
$P(E)$ is also very small.

Suppose, on the other hand, we attached the same low, prior probability
$\epsilon$ to the hypothesis, that the lottery is biased, but were sure
that if it is rigged in anyone's favour that someone is going to be Alice
(corresponding to a situation where we think the lottery organizer is
\emph{probably} honest, but happen to know that Alice is his wife). 
The prior probabilities now are $P(H_0)=1-\epsilon$ and $P(H_i) =
\epsilon \delta_{ia}$ for $i\ge 1$.  The posterior probability that the
lottery is fair, given that Alice won, is then
\begin{equation}
P(H_0|E) = \frac{1}{1+ N \epsilon/(1-\epsilon)} \, .
\label{eq:AliceLotteryBayes}
\end{equation}
If one took $\epsilon = 1/N$ this would give $P(H_0|E)\approx
1/2$---meaning that, even though we start out with a strong conviction
that the lottery organizer is probably honest, the event of his wife
winning makes us   very suspicious.  This also agrees with
ordinary intuition.

The paradox which argument 3 apparently represents can also be resolved
by analyzing the problem in Bayesian terms.  Let $H_p$ be the hypothesis
``tosses are independent and probability of heads is $p$ on every toss'',
and let
$E$ be the event that the particular sequence $\mathbf{s}$ containing $50$
heads and $50$ tails is obtained (as in argument~3).  Suppose we take
$P(H_p)=1$ for all $p$.  Then Eq.~(\ref{eq:bayes}) becomes
\begin{equation}
  P(H_p|E) = \frac{P(E|H_p)}{P(E)}
\end{equation}
where $P(E|H_p)=p^{50} (1-p)^{50}$ and $P(E)=\int_{0}^{1} P(E|H_p)
dp=9.8\times 10^{-32}$. Argument 3 depends on the fact that the likelihood
$P(E|H_{0.5}) = 7.9\times 10^{-31}$ is very small.  If one bases onself
on a supposed ``primitive'' or ``elemental''  resistance to accepting
highly improbable stories (Fisher~\cite{FisherStat}, p.~46) this means
that the occurrence of $E$ is grounds for rejecting $H_{0.5}$.  But if
one bases oneself on the posterior $P(H_p|E)$, as one logically should,
the paradox dissolves.  $P(E|H_p)$ and $P(E)$ are \emph{both} very small. 
Consequently, their ratio $P(H_p|E)$ is a well-behaved probability
density, with total area $=1$, and $96\%$ of its area concentrated in the
interval $(0.4,\, 0.6)$.

\section{No ``probable'' from an ``is''}
\label{sec:noprobable}
I have several times
mentioned Hume's argument for inductive scepticism. Hume is also
well-known for his fact-value distinction:  the principle that one cannot
validly infer an ``ought'' from an ``is''.  As he puts it~\cite{HumeA}
(p.~469):
\begin{quote}
  In every system of morality, which I have hitherto met with, I have
always remark'd, that the author proceeds for some time in the ordinary
way of reasoning \dots when of a sudden I am surpriz'd to
find, that instead of the usual copulations of propositions, \emph{is},
and \emph{is not}, I meet with no proposition that is not connected with
an \emph{ought}, or an \emph{ought not}.  This change is imperceptible;
but is, however, of  the last consequence.  For as this \emph{ought}, or
\emph{ought not}, expresses some new relation or affirmation, 'tis
necessary \dots
that a reason should be given, for what seems altogether inconceivable,
how this new relation can be a deduction from others, which are entirely
different from it.
\end{quote}
The point Hume is making here, that one cannot get a moral injunction
\emph{out}, at the end of an argument, unless one began by feeding a moral
assumption
\emph{in}, at the start, is nowadays usually taken for granted. However,
when Hume first advanced this proposition it seemed shocking.

Hume's point is that there exists no valid inference of the form
\begin{equation}
 F\Rightarrow N
\end{equation}
where  $F$ is a statement of fact and $N$ is a moral injunction.  If $N$
is validly inferred by an argument having $F$ among its premises, then the
inference must be of the form
\begin{equation}
M\wedge  F\Rightarrow N
\end{equation}
where $M$ is a prior moral assumption.

In Sections~\ref{sec:Bayes}--\ref{sec:orthodoxVbayes} we saw that
a similar principle applies to
 probabilistic reasoning.  A probability assignment $Q$ cannot be
inferred directly from a factual proposition $F$ without other 
input\footnote{
  Except in the trivial case, where one merely infers, from the
 fact that $x$ occurred, that $x$ was not impossible, or from the fact
 that $x$ did not occur, that $x$ was not certain.
  }.  
A retrodictive probabilistic inference must, instead, be of the form
(\emph{c.f.}\ Eq.~(\ref{eq:BayesGeneralForm}))
\begin{equation}
P\wedge  F\Rightarrow Q
\end{equation}
where $P$ is a prior probabilistic 
assumption.

In short:  one cannot get a ``probable'' from an ``is''.  
Probability statements are, in most respects, quite unlike moral
statements.  However, they have this logical feature in common.

\section{Predictive Inferences}
\label{sec:predictive}

Physical thinking has been much influenced by the idea that extremely
improbable events are FAPP (``for all practical purposes'') impossible. 
The idea is attractive because, by converting extreme probability
statements into statements of effective fact, it seems to circumvent the
need to give a clear physical interpretation of probability as such.

I have argued that
in a  retrodictive context this idea is unacceptable.  On the other
hand,  in predictive reasoning I think
 it (probably) is true that a sufficiently low probability event counts
as FAPP impossible.  However, one needs to be careful.

For instance, the
probability of Alice winning the lottery is $6\times 10^{-8}$.  It is
tempting to conclude that she has, from a predictive point of view,
effectively
\emph{no}  chance of
winning.  However, it will appear
on further reflection that matters are less straightforward.   It is true
that Alice will, if she is wise, take the event of her winning  to be
impossible FMPP (``for \emph{most} practical purposes'').
 She will
not, for example, make heavy financial commitments which she could not
 meet,  except in the event of her winning the lottery.  But if
she really did think  it to be impossible for \emph{all}
 practical purposes, she
would not have taken the practical step of buying a ticket in the first
place.

This example may appear frivolous.  So let us consider a less frivolous
one. The geological record suggests that the probability of the Earth 
being hit by a $10\, \mathrm{km}$ asteroid some time in the next $50$
years is
$\sim 10^{-6}$.  This is significantly greater than the probability of
Alice winning the lottery, but still rather small.  If we judged the
probability to be (say) $\sim 0.5$ then we might consider it worth
devoting a substantial fraction of the world's GDP to the problem of
trying to avert this potential catastrophe.  But as it is the probability
is small, and so we judge it more appropriate to expend most of the
world's resources on concerns that seem more pressing.  Nevertheless, the
US government does expend \emph{some} of its resources on the task of
tracking asteroids.  At least in the view of the US government, a chance
of $10^{-6}$  is not FAPP equivalent to a prediction, that the event in
question will certainly not happen.

However, a probability of $10^{-6}$  is still comparatively large.  It is
difficult to see how considerations of the kind just adduced could apply
to probabilities of $10^{-60}$, or  $10^{-600}$.  We do not, for instance,
consider it worth insuring against a  macroscopic violation of
the second law of thermodynamics.

 I am
therefore inclined to think that, once a probability is shrunk below a
certain point, the event in question does indeed count as FAPP impossible
so far as prediction is concerned.  This point is not very sharply
defined.   But it appears to me that a probability of $10^{-6}$  is rather
clearly on one side, and that a probability of  $10^{-60}$ is no less
clearly on the other.
However, that does not entirely settle the question.  We need to ask
where these numbers
$10^{-6}$ and
$10^{-60}$ are coming from.

I have been appealing to the idea of a fair bet, where one trades a small
stake for a potentially large gain, and a fair insurance, where one trades
a small premium for protection against a potentially large loss.  The
trade need not be conceived in financial terms.  This kind of reasoning
plays an essential role in medicine (where one has to balance the
debilitating effects of, say, chemotherapy  against the potential gain in
health), and  in theoretical research (where,
when choosing a project, one has to balance the labour to be expended
against the  intellectual value of the result, should the investigation
bear fruit). Indeed, I would say that, one way or another, it plays an
essential role in just about every department of life.

The conditions of human life constrain the size of any
appropriate stake or premium.  I am inclined to think that  the
significance of the numbers 
$10^{-6}$ and
$10^{-60}$  derives from  such facts as:
\begin{enumerate}
\item[] The  GDP of the world is $\sim 10^{16}$ US cents.
\item[] The age of the species \emph{Homo Sapiens} is $\sim 10^{13}$
seconds.
\item[] The volume of the solar system (out to the heliopause) is $\sim
10^{46}\,\mathrm{cm}^3$. 
\end{enumerate}
An alien species, which lived for times greatly in excess of $10^{10}$
years, and whose sphere of  interest extended over regions much
more than
$10^{10}$ light years across, might have very different ideas as to what
counts as FAPP impossible.

The concept of  something being impossible ``for all practical purposes''
is relative to the practical purposes of some \emph{particular} cognitive
agent.  So it is, in that sense,  subjective.

\section{Frequentism Revisited}
\label{sec:frequentism3}
The analysis in Sections~\ref{sec:Bayes}--\ref{sec:noprobable} shows that
probability assignments  cannot be reduced to
factual statements about the way things are in the world.  Probability
statements and factual statements have a fundamentally different logic.

In particular, the proposition
\begin{enumerate}
\item[$Q$:] ``The probability of this coin coming up heads is $0.5$''
\end{enumerate}  is not, as finite frequentists think, FAPP equivalent to
any factual proposition concerning the number of heads that will occur in
a sufficiently long sequence of tosses.

Furthermore, $Q$ cannot, as falsificationists like Popper think, be FAPP
falsified by any factual proposition concerning the number of heads that
will occur in a sufficiently long sequence of tosses.

The bare fact
\begin{enumerate}
\item[$F$:] ``The coin came up heads on each of $1000$ successive tosses''
\end{enumerate}
is no more reason for thinking that $Q$ is false than the
bare fact
\begin{enumerate}
\item[$F'$:] ``The coin came up heads on $500$ out of $1000$ successive
tosses''
\end{enumerate}

Of course, if a coin did in reality come up heads on each of $1000$
successive tosses, one would in practice take that as very strong evidence
that the coin was biased.  However, this conclusion would be based, not on
the bare fact $F$, but on the dressed fact $P\wedge F$, where $P$ is a
prior probabilistic assumption.

We have become so habituated to the frequentist way of thinking that this
assertion may appear paradoxical.  But in fact the point, that $F$ is
perfectly consistent with the coin being fair, is  something that is
taught in every elementary textbook.

As is well-known, gamblers are prone to believe that if a coin has come
up heads on (say) $2$ successive tosses, then the probability of
heads on the next toss is $<1/2$.  One of the first things students are
taught is that this is a fallacy.  Provided the tosses are independent,
the probability of heads on the next toss continues to be $1/2$,
irrespective of how many times heads has come up on the preceding
tosses.  

This statement is an elementary consequence of the assumption of
independence.
It is, of course, true that in practice hardly anyone would continue to
believe that a coin is fair if it kept on coming up heads in toss after
toss.  But that only shows that, in practice,  hardly anyone   would
seriously believe that the tosses are 
independent\footnote{
  I should say that there is an ambiguity here, if one looks
  at it in von Mises's objectivist terms.  As von Mises sees it, there
  is a true probability of heads $p$.  If we knew the value 
  of $p$ the probability of obtaining a sequence $\mathbf{s}$
  would be given by the conditional  distribution
$P(\mathbf{s}|p)$ defined by Eq.~(\ref{eq:Bayes1}).  In that case the
probability of heads on the $n^{\mathrm{th}}$ toss \emph{is} independent
of what occurred on the preceding tosses.  However, if we do \emph{not}
know the value of $p$ then the probabilities have to be calculated 
using the unconditioned distribution $P(\mathbf{s})=\int_{0}^{1}
P(\mathbf{s}|p) f_{\mathrm{i}}(p) dp$.  In that case the tosses are 
typically
\emph{not} independent ---as can be seen from Eq.~(\ref{eq:antiFreqC})
below.
}.

In practice most people would start out with the belief that the
probability of heads on any given toss $\approx 1/2$.  But they would
\emph{also} start out with the belief  that if, for example, heads
should occur on each of the first $n$ tosses, for some large number
$n$, then that would mean that the probability of heads on the
$(n+1)^{\mathrm{th}}$ toss
$\approx 1$. Symbolically:
\begin{equation}
 P(E_r)  \approx 1/2 \qquad \text{for all $r$}
\label{eq:antiFreqA}
\end{equation}
and
\begin{equation}
P(E_{n+1}|E_1\wedge \dots \wedge E_{n})  \approx 1
\label{eq:antiFreqB}
\end{equation} 
where $E_r$ is the event ``heads on the $r^{\mathrm{th}}$ toss''.  That
pair of propositions is inconsistent with the proposition, that
successive tosses are independent.

The logical basis for these intuitive assumptions is best understood by
looking at the problem in Bayesian terms.  From
Eq.~(\ref{eq:Bayes2}) the prior probability of obtaining, in the first $n$
tosses, a sequence
$\mathbf{s}$ containing $h(\mathbf{s})$ heads  is
\begin{equation}
P(\mathbf{s}) = \int_{0}^{1} p^{h(\mathbf{s})} (1-p)^{n-h(\mathbf{s})}
f_{\mathrm{i}} (p) dp \, .
\end{equation}
If  the prior distribution $f_{\mathrm{i}}$ is symmetric
about the mid-point $p=1/2$
\begin{equation}
 P(E_r) = \int_{0}^{1} p f_{\mathrm{i}}(p) dp = 1/2 \qquad \text{for all
$r$}
\end{equation}
in agreement with Eq.~(\ref{eq:antiFreqA}).  Suppose it is also the case
that (for example)
$f_{\mathrm{i}}$ is continuous and $f_{\mathrm{i}}(1)> 0$.  Then
\begin{equation}
 P(E_1 \wedge \dots \wedge E_n) = \int_{0}^{1} p^n f_{\mathrm{i}}(p) dp 
\approx f_{\mathrm{i}}(1)/(n+1)
\end{equation}
for suffiently large $n$.  Consequently
\begin{equation}
 P(E_{n+1}|E_1\wedge \dots \wedge E_n) \approx \frac{n+1}{n+2} \approx 1
\label{eq:antiFreqC}
\end{equation}
for sufficiently large $n$---in agreement with Eq.~(\ref{eq:antiFreqB}).

However, the value of $n$ at which $P(E_{n+1}|E_1\wedge \dots \wedge
E_n)$ becomes close to $1$ will depend on the choice of
$f_{\mathrm{i}}$.  For some choices of $f_{\mathrm{i}}$ a comparatively
short run of heads will be enough to convince us that the coin is
probably biased.  For other choices a much longer run will be needed.  And
if $f_{\mathrm{i}} = \delta (p-0.5)$ then nothing will convince us.  For
that choice of $f_{\mathrm{i}}$
\begin{equation}
 P(E_{n+1}|E_1\wedge \dots \wedge E_n) = P(E_{n+1}) = 1/2
\end{equation}  
for all $n$.  So we would continue to believe that the probability of
heads on the next toss is $1/2$ even if (\emph{per impossibile}) the coin
had come up heads on each of the preceding $10^{100}$ tosses.

The distribution $f_{\mathrm{i}}$ represents our background assumptions. 
In other words, it represents the  probabilistic context.   The
inference is  sensitive to that context.   For some
 contexts
$P$ the conjunction 
$P\wedge \left(\bigwedge_{i=1}^{10^3} E_i\right) $ implies that the coin
will almost certainly come up heads on the next toss.  For others it
implies that the probability of this happening $\approx 1/2$.   
And if one considers the fact $\bigwedge_{i=1}^{10^3} E_i$ in isolation,
devoid of any context, then no conclusion is possible.  A \emph{bare}
fact has no (interesting) probabilistic implications.

It is the contextuality of retrodictive probabilistic inferences which
defeats the frequentist programme.

Finally, let us note that on this account of the matter nothing is
falsified.  The proposition $P(E_{n+1}|E_1\wedge \dots \wedge E_{n}) 
\approx 1$ is built into our starting assumptions.  So if, after 
$E_1\wedge \dots \wedge E_{n}$ has actually occurred, we then believe
that $E_{n+1}$ is close to certain, we are not believing anything that
contradicts our starting assumptions.

\section{Probability is Epistemic}
\label{sec:normative}
Before the outcome is known we consider that Alice is most unlikely to win
the lottery.  But then she does win.  We do not conclude that we were
wrong.  Instead, we conclude that Alice was very lucky.  So what, exactly,
is meant by saying that Alice is most unlikely to win the lottery?

The frequentist strategy, of replacing a  single lottery draw with a
long sequence of draws, does nothing to answer that question.  All
 it achieves is to replace the statement, that something is very
unlikely, with the statement, that something else is even more unlikely.
It does not take us any nearer to understanding what it actually
\emph{means} to say that something is very unlikely.

Nevertheless, probability
assignments are clearly not devoid of content. 
To see what
their content is, consider the proposition ``Alice was very lucky to win
the lottery'' asserted after it is known that she did in fact win the
lottery.   This means something like:  though Alice \emph{did} win
 the lottery,
she could not reasonably have \emph{expected} to win.  
The proposition is not
unrelated to  empirical facts about the lottery.  But the
primary focus is on Alice, and her subjective attitude. 

Suppose Alice confidently
asserts ``I am going to win the lottery'', before the result is known.  If
she then wins the lottery, we have to agree that her statement was
factually correct.  But we do not have to agree that her confidence was 
justified.  And that is the important sense in which one can meaningfully
describe something as very unlikely, even though it has actually
happened. 

Compare: 
\begin{quote}
\textbf{Case
1:} Alice neglects to make any pension contributions on the grounds that,
when the time comes, she can always buy a lottery ticket.  When she
reaches the age of 65 she buys one lottery ticket, the ticket wins, and
she enjoys a comfortable retirement.
\end{quote}
\begin{quote}
 \textbf{Case 2:} Bob neglects to make any
pension contributions on the grounds that he has \pounds 10,000,000
invested in several large companies listed on the  Stock Exchange. 
Shortly before he retires all the companies go bankrupt.  He lives out his
remaining years in penury. 
\end{quote}
Even though it is Alice's expectations that
are actually fulfilled, we would still (most of us) consider that Bob's
expectations were more reasonable.

The statement, that Alice's behaviour is unreasonable, expresses
 some
kind of value judgment.  De Finetti would doubtless object to the word
``unreasonable''.  He would prefer to describe Alice's behaviour as
``crazy'' (de Finetti~\cite{Fin2}, p.~175).  However, he would still be
expressing a negative 
evaluation.

In this paper I have been making the negative point: 
that probability statements do not reduce to purely factual statements,
concerning events out there in the world.   I remain very
uncertain regarding the positive question, as to how probability
statements should be interpreted. 

However, although I have no definite
opinion about many of the details, I feel that in general terms
Maxwell must be right when he says that the theory of probability has an
essentially normative significance.  We use it to evaluate alternative
beliefs and proposed courses of action:  to decide what, in given
circumstances, it is best or most appropriate to think and do.

\section{Reality:  Einsteinian or Bohrian?}
\label{sec:conclusion}
For  a long time Einstein strongly objected to the indeterminism of
quantum mechanics.  As he put it in a letter to Born~\cite{Born} (p.~91),
written in 1926:
\begin{quote}
 Quantum mechanics is certainly imposing.  But an inner voice tells me
that it is not yet the real thing.  The theory says a lot, but does not
really bring us any closer to the secret of the `old one'.  I, at any
rate, am convinced that \emph{He} is not playing at dice.
\end{quote}
He expressed the same view in another letter (\emph{ibid} p.~149) written
as late as 1944.  I think people often find it difficult to understand
why Einstein was  so emphatic in his rejection of
a dice-playing God.  Quantum mechanics presents many obstacles to the
understanding. But the concept of an objective chance  seems
intuitively very natural.  The commonsense world is full of entities
endowed with propensities of one kind or another.    At least as judged
by the standards of commonsense, if anything is paradoxical, it is the
rigid determinism of classical physics.  

Einstein apparently came to feel this himself in the end.  In 1954
Pauli~\cite{Born} (p.~221) reports him as ``\emph{disput[ing]} that he
uses as criterion for the admissibility of a theory the question:  `Is it
rigorously deterministic?'\,''.  
However, it seems to me that Einstein gave in too easily.  There are 
very strong objections to the dice-playing aspect of quantum mechanics,
if one approaches it from Einstein's philosophical standpoint.

The commonsense world is indeed full of real chances.  But it is also full
of other things, such as real colour \emph{qualia}.
It has been accepted since the $17^{\mathrm{th}}$ century that colour
\emph{qualia} are not in fact physically real at all.  According to
Newton~\cite{Newton} (p.~124--5):
\begin{quote}
``\dots the Rays to speak properly are not coloured.  In them there is
nothing else than a certain Power and Disposition to stir up a Sensation
of this or that Colour''
\end{quote} 
and
\begin{quote}
``\dots Colours in the Object are nothing but a Disposition to reflect
this or that sort of Rays more copiously than the rest; in the Rays they
are  nothing but their Dispositions to propagate this or that Motion into
the Sensorium, and in the Sensorium they are Sensations of those Motions
under the Forms of Colours''
\end{quote}
In short:  colour \emph{qualia} are mental, not physical.  

Interestingly Newton makes no attempt to justify this assertion, either on
experimental, or on any other grounds.  Instead he proposes it as a
\emph{definition}:  something we should just accept.
However, I think it is worth asking why.

The obvious answer to this question is that real \emph{qualia} 
would not fit in with the kind of mechanical picture which Newton is
trying to construct.  However, I want to focus on a different point: 
namely, that perceived colour \emph{qualia} clearly depend on
physiological peculiarities of the  human eye-brain
system.  For instance, there are conventionally said to be $7$ colours in
the visible spectrum.  The number $7$ is, perhaps, a little arbitrary. 
But I believe no one experiences, say,
$10^6$ distinct
\emph{qualia}.  The explanation for this must presumably lie with
properties of the human eye and brain---not with properties of the
electromagnetic  spectrum.

It follows that, if one were to construct a theory in which the
\emph{qualia} were represented  as
physically real, then  one would be building
into one's picture of physical reality features which actually derive
from ourselves.  

Newton's standpoint, and Einstein's standpoint in his later years, was
that the aim of physics is to construct a  a
picture of things as they are intrinsically, in themselves, without any
trace of subjective contamination.   From that point of view the concept
of a real colour
\emph{qualium} is  unacceptable.

If that is your point of view, then the concept of a real probability,
or propensity, must be equally unacceptable.   According to the analysis
in the preceding sections a probability statement is a statement about
what, in given circumstances, one may reasonably think or do.  This
reference to our own cognitive processes means that the concept can have
no place in the purely objective world-view which Einstein was trying to
construct.  

In short, it seems to me that Einstein, given his conceptual standpoint,
had every reason to reject the notion of a dice-playing God.  If one
chooses to follow the Einsteinian road then one had better look for a
fully causal interpretation of quantum mechanics, such as the de
Broglie-Bohm interpretation.

However, there is another possibility.  One could, instead, choose to go
down what might be called the Bohrian road.  Nothing with quantum
mechanics built into it could be called commonsensical.  But a  Bohrian
view of the world would be like commonsense in as much as it would make
essential reference to the cognitive agent, whose view it is.  It would
be colour-full and value-laden.

As
it stands now the Bohrian view is, to my mind, very obscure.  But it may
have within it the potential for fruitful development.

\subsubsection*{Acknowledgements}  I am grateful to C.A.~Fuchs (who
originally excited my interest in this question) and to
H.~Brown, P.~Busch, J.~Butterfield,  M.~Donald,  L.~Hardy, T.~Konrad,
C.~Timpson and  J.~Uffink for stimulating discussions.


\begin{thebibliography}{99}
\bibitem{FuchsEtAl} C.M.~Caves, C.A.~Fuchs, and R.~Schack, ``Quantum
Probabilities as Bayesian Probabilities'', \emph{Phys. Rev. A}\
\textbf{65}, 022305 (2002).
\bibitem{FuchsEtAlB}  C.M.~Caves, C.A.~Fuchs, and R.~Schack, ``Unknown
Quantum States:  the Quantum de Finetti Representation'',
\emph{J.~Math.~Phys.}\ \textbf{43}, 4537 (2002).
\bibitem{FuchsEtAlC} C.M.~Caves, C.A.~Fuchs, and R.~Schack, ``Conditions
for Compatibility of Quantum-State Assignments'', 
\emph{Phys.\ Rev.\ A}\ \textbf{66}, 062111 (2002).
\bibitem{FuchsSam1} C.A.~Fuchs ``Notes on a Paulian Idea'', e-print
quant-ph/0105039.
\bibitem{FuchsSam2} C.A.~Fuchs, ``Quantum Mechanics as Quantum
Information (and only a little more)'', e-print quant-ph/0205039.
\bibitem{CushingA} J.T.~Cushing, \emph{Quantum Mechanics:  Historical
Contingency and the Copenhagen Hegemony} (The University of Chicago
Press, Chicago, 1994).
\bibitem{BallEd} Editorial Comment, \emph{Rev.\ Mod.\ Phys.}\
\textbf{42}, 357 (1970).
\bibitem{BallPaper}  L.E.~Ballentine, ``The Statistical Interpretation of
Quantum Mechanics'', \emph{Rev.\ Mod.\ Phys.}\
\textbf{42}, 358 (1970).
\bibitem{Bell} J.S.~Bell, \emph{Speakable and Unspeakable in Quantum
Mechanics} (Cambridge University Press, Cambridge,1987).
\bibitem{HardyA} L.~Hardy, ``Quantum Theory from Five Reasonable
Axioms'', e-print quant-ph/0101012.
\bibitem{HardyB} L.~Hardy, ``Why Quantum Theory?'', in J.~Butterfield and
T.Placek (eds), 
\emph{Proceedings of the NATO Advanced Research Workshop on Modality,
Probability and Bell's Theorem} (IOS Press, Amsterdam, 2002);  e-print
quant-ph/0111068.
\bibitem{Pitowsky} I.~Pitowsky, ``Betting on the Outcomes of Measurements: 
A Bayesian Theory of Quantum Probability'', \emph{Stud.\ Hist.\ Phil.\
Mod.\ Phys.}\ \textbf{34}, 395 (2003).
\bibitem{Perey} F.G.~Perey, ``Probabilities as Measures of Information'',
e-print quant-ph/0310073.
\bibitem{ValentiniA} A.~Valentini, ``Hidden Variables, Statistical
Mechanics and the Early Universe'' in J.~Bricmont \emph{et al},
\emph{Chance in Physics:  Foundations and Perspectives} (Springer,
Berlin, 2001).
\bibitem{ValentiniB} A.~Valentini, ``Signal-Locality in
Hidden-Variables Theories'', \emph{Phys.\ Lett.\ A} \textbf{297}, 273
(2002).
\bibitem{HackingA} I.~Hacking, \emph{The Emergence of Probability}
(Cambridge University Press, Cambridge, 1975).
\bibitem{Daston} L.~Daston, \emph{Classical Probability in the
Enlightenment} (Princeton University Press, Princeton, 1988).
\bibitem{Poincare} H.~Poincar\'{e}, \emph{Science and Hypothesis} (Dover,
New York, 1952).  English translation, first published 1905.
\bibitem{GilliesA}  D.~Gillies, \emph{Philosophical Theories of
Probability} (Routledge, London, 2000).
\bibitem{vonPlato} J.~von Plato, \emph{Creating Modern Probability}
(Cambridge University Press, Cambridge, 1994).
\bibitem{SklarA} L.~Sklar, \emph{Physics and Chance} (Cambridge
University Press, Cambridge, 1993).
\bibitem{SklarB} L.~Sklar (ed.), \emph{Probability and Confirmation}
(Garland Publishing, New York, 2000).
\bibitem{Guttman} Y.M.~Guttman, \emph{The Concept of Probability in
Statistical Physics} (Cambridge, Cambridge University Press, 1999).
\bibitem{JeffreysB} H.~Jeffreys, \emph{Theory of Probability},
$3^{\mathrm{rd}}$ edition  (Clarendon Press, Oxford, 1961).
\bibitem{Laplace} P.S.~de Laplace (trans. F.W.~Truscott and F.L.~Emory),
\emph{A Philosophical Essay on Probabilities}  (Dover, New York, 1951).
French original published 1820.
\bibitem{Keynes} J.M.~Keynes, \emph{A Treatise on Probability}
(Macmillan, London, 1921).
\bibitem{JeffreysC} H.~Jeffreys, \emph{Scientific Inference},
$3^{\mathrm{rd}}$ edition (Cambridge University Press, Cambridge, 1973).
\bibitem{Carnap} R.~Carnap, \emph{Logical Foundations of Probability}
(University of Chicago Press, Chicago, 1962).
\bibitem{Lewis} D.~Lewis, \emph{Philosophical Papers}, vol.~2 (Oxford
University  Press, Oxford,  1986).
\bibitem{Fin2} B.~de Finetti,  ``Probabilism'', English Translation,
\emph{Erkenntnis} \textbf{31}, 169 (1989).  Italian original published
1931.
\bibitem{Fin1} B.~de Finetti (trans.\ A.~Mach\'{i} and A.~Smith),
\emph{Theory of Probability} (Wiley, New York, 1975).  Italian original
published 1971.
\bibitem{Savage} L.J.~Savage, \emph{The Foundations of Statistics},
$2^{\mathrm{nd}}$ edition (Dover, New York, 1972).
\bibitem{Bernardo} J.M.~Bernardo and A.F.M.~Smith, \emph{Bayesian Theory}
(John Wiley and Sons, Chichester, 1994).
\bibitem{Jaynes} E.T.~Jaynes (ed. R.D.~Rosenkrantz), \emph{Papers on
Probability, Statistics and Statistical Physics} (Reidel, Dordrecth,
1983).
\bibitem{Howson} C.~Howson and P.~Urbach, \emph{Scientific Reasoning: 
the Bayesian Approach} (Open Court, La Salle, 1989).
\bibitem{Earman}  J.~Earman, \emph{Bayes or Bust? A Critical Examination
of Bayesian Confirmation Theory} (MIT Press, Cambridge Mass, 1992).
\bibitem{Venn}  J.~Venn, \emph{The Logic of Chance}, $4^{\mathrm{th}}$
edition (Chelsea, New York, 1962).  Reprint of $3^{\mathrm{rd}}$
edition,  published 1888. 
\bibitem{Mises} R.~von Mises, \emph{Probability, Statistics and Truth}
(Dover, New York, 1981).  Reprint of $2^{\mathrm{nd}}$ revised English
edition,
 published 1957.
\bibitem{MisesB} R.~von Mises (ed.\ H.~Geiringer),
\emph{Mathematical Theory of Probability and Statistics} (Academic Press,
New York, 1964).
\bibitem{Reich} H.~Reichenbach, \emph{The Theory of Probability}
(University of California Press, Berkelely, 1971).
\bibitem{PopperB} K.R.~Popper, \emph{The Logic of Scientific Discovery}
(Hutchinson, London, 1959).
\bibitem{Fraassen} B.C.~van Fraassen, \emph{The Scientific Image}
(Clarendon Press, Oxford, 1980).
\bibitem{Ramsey}  F.P.~Ramsey, ``Truth and Probability'' reprinted in
Sklar~\cite{SklarB}.  First published 1931.
\bibitem{Fisher}  R.A.~Fisher (ed. J.H.~Bennett), \emph{Statistical
Methods, Experimental Design, and Scientific Inference} (Oxford
University Press, Oxford, 1990).
\bibitem{Neyman}  J.~Neyman and E.S.~Pearson, \emph{Joint Statistical
Papers} (Cambridge University Press, Cambridge, 1967).
\bibitem{JaynesB} E.T.~Jaynes, ``Confidence Intervals vs Bayesian
Intervals'' in W.L.~Harper and C.A.~Hooker (eds) \emph{Foundations of
Probability Theory, Statistical Inference, and Statistical Theories of
Science} (Reidel, Dordrecht, 1976).  Page references to version reprinted
in Jaynes~\cite{Jaynes}.
\bibitem{HumeA} D.~Hume (ed.~L.A.~Selby-Bigge, revised P.H.~Nidditch),
\emph{A Treatise of Human Nature}, $2^{\mathrm{nd}}$ edition (Clarendon
Press, Oxford, 1978).  Originally published 1739-40.
\bibitem{HumeB} D.~Hume (ed.~L.A.~Selby-Bigge, revised P.H.~Nidditch),
\emph{Enquiries Concerning Human Understanding and Concerning the
Principles of Morals}, $3^{\mathrm{rd}}$ edition (Clarendon Press, Oxford,
1975).  Originally published 1777.
\bibitem{Einstein} P.A.~Schilpp, \emph{Albert Einstein: Philosopher
Scientist}, $3^{\mathrm{rd}}$ edition (Open Court, La Salle, 1982).
\bibitem{Hajek} A.\ H\'{a}jek, ``\,`Mises Redux'---Redux: fifteen
arguments against finite frequentism'', 
\emph{Erkenntnis} \textbf{45},
209--227 (1997).
\bibitem{Feller} W.~Feller, \emph{An Introduction to 
Probability Theory
and its Applications} (Wiley, New York, 1950).
\bibitem{Jeffrey} R.C.~Jeffrey, ``Mises Redux'' in R.E.~Butts and
J.~Hintikka (eds), \emph{Basic Problems in Methodology and Linguistics: 
$5^{\mathrm{th}}$ International Congress of Logic, Methodology, and
Philosophy of Science pt.~3} (Reidel, Dordrecht, 1977).
\bibitem{GilliesB} D.A.~Gillies, \emph{An Objective Theory of
Probability} (Methuen, London, 1973).
\bibitem{Good} N.~Goodman, \emph{Fact, Fiction and Forecast}  
(Harvard
University Press, Cambridge Mass, 1955).
\bibitem{PopperC} K.R.~Popper, ``The Propensity Interpretation of
Probability'', \emph{Brit.\ J.\ Phil.\ Sci.}\ \textbf{10}, 25--42
(1959).
\bibitem{PopperD} K.R.~Popper, \emph{Realism and the Aim of Science}
(Hutchinson, London, 1983). 
\bibitem{PopperE} K.R.~Popper, \emph{A World of Propensities} (Thoemmes,
Bristol, 1990).
\bibitem{WittgensteinA} L.~Wittgenstein (ed.~R.~Rhees, trans.~A.~Kenny),
\emph{Philosophical Grammar} (Basil Blackwell, Oxford, 1974).
\bibitem{WittgensteinB} L.~Wittgenstein (ed.~R.~Rhees,
trans.~R.~Hargreaves and R.~White),
\emph{Philosophical Remarks} (Basil Blackwell, Oxford, 1975).
\bibitem{WittgensteinC} L.~Wittgenstein (trans.~G.E.M.~Anscombe),
\emph{Philosophical Investigations}, $3^{\mathrm{rd}}$ edition (Basil
Blackwell, Oxford, 1968).
\bibitem{FisherStat} Fisher, \emph{Statistical Methods and Scientific
Inference}, $3^{\mathrm{rd}}$ edition.  Page references to version
reprinted in Fisher~\cite{Fisher}.
\bibitem{FisherDesign} R.A.~Fisher, \emph{The Design of Experiments},
$8^{\mathrm{th}}$ edition.  Page references to version reprinted in
Fisher~\cite{Fisher}.
\bibitem{Bohm}  D.~Bohm and B.J.~Hiley, \emph{The Undivided Universe}
(Routledge, London, 1993).
\bibitem{Holland}  P.R.~Holland, \emph{The Quantum Theory of Motion}
(Cambridge University Press, Cambridge, 1993).
\bibitem{Born} M.~Born and A.~Einstein (trans.~I.~Born), \emph{The
Born-Einstein  Letters} (Macmillan, London, 1971).
\bibitem{Newton} I.~Newton, \emph{Opticks}, based on the
$4^{\mathrm{th}}$ edition, 1730 (Dover, New York, 1952).
\end{thebibliography}
\end{document}